\documentclass [12pt]{article}
\usepackage[round]{natbib}

\usepackage {graphicx}
\usepackage{epsfig,amsmath,latexsym,amssymb}
\usepackage{bm}
\usepackage{caption}

\usepackage{booktabs}
\usepackage{multirow}
\newtheorem{algorithm_new}{Algorithm}[section]

\begin{document}

\title{\vspace{0cm} A unified pricing of variable annuity guarantees under the optimal stochastic control framework}

\author{\textnormal{Pavel V.~Shevchenko$^{1}$ and Xiaolin Luo$^{2}$}}

\date{\footnotesize{Draft paper: 16 April 2016}}

\maketitle

\begin{center}\vspace{-0.5cm}
\footnotesize { \textit{$^{1}$ CSIRO Australia, e-mail: Pavel.Shevchenko@csiro.au \\
$^{2}$ CSIRO Australia, e-mail: Xiaolin.Luo@csiro.au
} }
\end{center}

\begin{abstract}{\footnotesize
\noindent In this paper, we review pricing of variable annuity
living and death guarantees offered to retail investors in many countries. Investors purchase these products to
take advantage of market growth and protect savings. We present
pricing of these products via an optimal stochastic control
framework, and review the existing numerical methods. For numerical
valuation of these contracts, we develop a direct integration method
based on Gauss-Hermite quadrature with a one-dimensional cubic
spline for calculation of the expected contract value, and a
bi-cubic spline interpolation for applying the jump conditions
across the contract cashflow event times. This method is very efficient when compared
to the partial differential equation methods if the transition
density (or its moments) of the risky asset underlying the contract is
known in closed form between the event times. We also present
accurate numerical results for pricing of a Guaranteed Minimum
Accumulation Benefit (GMAB) guarantee available on the market that can
serve as a benchmark for practitioners and researchers developing
pricing of variable annuity guarantees.

}


\vspace{0.5cm} \noindent \textbf{Keywords:} \emph{variable annuity,
guaranteed living and death benefits, guaranteed minimum
accumulation benefit, optimal stochastic control, direct integration method.}

\end{abstract}

\pagebreak

\section{Introduction}
\label{sec:introduction} Many wealth management and insurance companies
worldwide are offering investment products known as \emph{variable annuities} (VA) with some guarantees of living and death benefits to
assist investors with managing pre-retirement and post-retirement plans. These products take advantage of market growth while
 provide protection of the savings against market downturns. Insurers started to offer
these products from the 1990s
in United States. Later, these products became popular in Europe, UK
and Japan and more recently in Australia.
The VA contract cashflows received by the policyholder are linked to the investment portfolio choice and performance (e.g. the choice of mutual fund and its strategy) while traditional annuities provide a pre-defined income stream in exchange for the lump sum payment.
According to LIMRA (Life Insurance and Market Research Association) reports, the VA market is huge: VA sales in United States were \$158 billion in 2011, \$147 billion in 2012 and \$145 billion in 2013.

The types of VA guarantees (referred in the literature as \emph{VA riders}) offered for investment portfolios are classified as
 \emph{guaranteed minimum withdrawal benefit} (GMWB), \emph{guaranteed minimum accumulation benefit} (GMAB), \emph{guaranteed minimum income benefit}
 (GMIB) and \emph{guaranteed minimum death benefit} (GMDB). These guarantees, generically denoted as GMxB, provide different types
 of protection against market downturns and policyholder death. GMWB allows withdrawing funds from the VA account up to some pre-defined
  limit regardless of investment performance during the contract; GMAB and GMIB both provide a guaranteed investment account
  balance at the contract maturity that can be taken as a lump sum or standard annuity respectively. \emph{Guaranteed lifelong withdrawal benefit} (GLWB), a specific type of GMWB, allows  withdrawing funds at the contractual rate as long as the policyholder
  is alive.
  GMDB provides a specified payment if the policyholder dies.
Precise specifications of the products within each type can vary across companies and some products may include combinations of these guarantees.

A good overview of VA products and the development
of their market can be found in \cite{bauer2008universal}, \cite{Ledlie2008variableannuities} and \cite{Kalberer2009}. There have been a number of papers in academic literature considering pricing of these products. Most of these are focused on pricing VA riders under the pre-determined (\emph{static}) policyholder behaviour in withdrawal and surrender. Some studies include pricing under the active (\emph{dynamic}) strategy when the policyholder `\emph{optimally}' decides the amount of withdrawal at each withdrawal date depending on the information available at that date.
Standard Monte Carlo (MC) method can easily be used to estimate price  in the case of pre-defined withdrawal strategy but handling
the dynamic strategy requires backward in time solution that can be done only via the partial differential equation (PDE),
direct integration or regression type MC methods.

In brief, pricing under the static and dynamic withdrawal strategies
via PDE based methods has been developed in
\cite{milevsky2006financial}, \cite{dai2008guaranteed} and
\cite{Forsyth2008}. \cite{bauer2008universal} develops a unified
approach with numerical estimation via MC and direct integration
methods. The direct integration method was developed further in
\cite{LuoShevchenkoGMWB2015, LuoShevchenkoGMWDB2015} using
Gauss-Hermite quadrature and cubic interpolations.
\cite{bacinello2011unifying} consider many VA riders under
stochastic interest rate and stochastic volatility if the
policyholder withdraws at the pre-defined contractual rate or
completely surrenders the contract.  Their pricing is accomplished
either by the ordinary MC or Least-Squares MC to account for the
optimal surrender. Typically, pricing of VA riders is considered under
the assumption of geometric Brownian motion for the risky asset underlying the contract,
though a few papers looked at extensions such as stochastic interest
rate and/or stochastic volatility, see e.g. \cite{ForsythVetzal2014},
\cite{LuoShevchenko2016GMWBstInterest},
\cite{bacinello2011unifying}, \cite{huang2015}.

\cite{Azimzadeh2014} prove the existence of an optimal \emph{bang-bang} control
for GLWB contract when the
contract holder can maximize contract writer's losses by only ever performing non-withdrawal,
withdrawal at the contract rate or full surrender. However, they also demonstrate that the related GMWB contract does
not satisfy the bang-bang principle other than in certain degenerate cases. \cite{huang2015} developed a regression-based
MC method for pricing GLWB under the bang-bang strategy in the case of stochastic volatility. GMWB pricing under the bang-bang strategy
 was studied in \cite{LuoShevchenko2015surrender}. The difficulty with applying the well known Least-Squares MC introduced in \cite{Longstaff2001}
 for pricing VA riders under the optimal strategy is due to the fact that the paths of the underlying VA wealth account are affected by the
 withdrawals. In principle, one can apply \emph{control randomization} methods extending Least-Squares MC to handle optimal
  stochastic control problems with controlled Markov processes
recently developed in \cite{Kharroubi2014},  but the accuracy and
robustness of this method for pricing  VA riders have not been
studied yet.

One common observation in the above mentioned literature is that
pricing under the optimal strategy often leads to prices
significantly higher than   observed on the market. These studies
rely on the option pricing risk-neutral methodology in quantitative finance to
find a \emph{fair fee}.  Here, the fundamental idea is
to find the cost of a dynamic self-financing replicating
portfolio which is designed to provide an amount at least equal to
the payoff of the contract. The cost of establishing this
hedging strategy is the no-arbitrage price of the contract. This is
under the assumption that the contract holder adopts an optimal strategy (exercise
strategy maximising the monetary value of the contract). If the
purchaser follows any other exercise strategy, the contract writer
will generate a guaranteed profit if continuous hedging is performed. Of course the strategy optimal  in this
sense is not related to the policyholder circumstances. In pricing VA with guarantees, it is
reasonable to consider alternative assumptions regarding the
investor's withdrawal strategy. This is because an investor may
follow what appears to be a \emph{sub-optimal} strategy that does not
maximise the monetary value of the option. This could be
due to reasons such as liquidity needs, tax
and other personal circumstances. Moreover, mortality risk is diversified by the
contract issuer through selling many contracts to many people while
the policyholder cannot do it. Also, there might be no liquid
secondary market for  VAs on which the policy could be sold (or
repurchased) at its fair value.
The
policyholder may act optimally with respect to his preferences and
circumstances but it may be different from the optimal strategy
that maximises the monetary value of the contract.
In this case we calculate a fair fee to be deducted
in order to finance a dynamic replicating portfolio for the guarantees
(options) embedded in the contract under the assumption of a
particular exercise strategy. The replicating portfolio will
 provide sufficient funds to meet any future payouts that arise
from writing the contract.

However, the fair fee obtained under the assumption that
investors behave optimally to maximise the value of the guarantee does offer
an important benchmark because it is a worst case scenario for the
contract writer. Also, as noted in
\cite{hilpert2014effect}, secondary markets for equity linked
insurance products (where the policyholder can sell their contracts)
are growing. Thus,  third parties can potentially generate
guaranteed profit through hedging strategies from financial products
such as VA riders which are not priced under the  assumption of
 the optimal withdrawal strategy. \cite{knoller2015propensity} mentions
several companies recently suffering large losses related to
increased surrender rates, indicating that either charged fees were not
sufficiently large or that hedging program did not perform as expected.

 One way to analyze the withdrawal behavior of VA holder and evaluate the need of these products is
 to solve the life-cycle utility model accounting for consumption, housing, bequest and other real life circumstances.
 Developing a full life-cycle model with all preferences and required parameters is challenging  but there are already several
  contributions reporting some interesting findings in this direction: \cite{moenig2012optimal,horneff2015optimal, gao2012optimal,steinorth2015valuing}. This topic will not be considered in this paper. It is also important to note a recent
  paper by \cite{bauer2015behavior} considering the pricing under the optimal strategy in the presence of taxes via \emph{subjective}
  risk-neutral valuation methodology. They demonstrated that including taxes significantly affects the value of the VA  withdrawal
  guarantees  producing results in line with empirical market prices.


In this paper we review pricing of living and death benefit guarantees offered with VAs, and present a unified optimal stochastic control framework for pricing these contracts. The main ideas have been developed and appeared in some forms in a number of other papers. However, we believe that our
presentation is easier to understand and implement.  We also present
direct integration method based on computing the expected contract
values in a backward time-stepping through
a high order Gauss-Hermite integration quadrature applied on a cubic
spline interpolation. This method can be applied when transition
density of the underlying asset between the contract cashflow event dates \emph{or
  its moments} are known in closed form. We have  used this for pricing specific financial derivatives and some
   simple versions of VA guarantees in \cite{LuoShevchenkoGHQC2014, LuoShevchenkoGMWB2015}. Here, we adapt and extend the
   method to handle pricing VA riders in general.
As a numerical example, we calculate accurate prices of GMAB with possible annual \emph{ratchets} (reset of the guaranteed capital
 to the investment portfolio  value if the latter is larger on anniversary dates) and allowing optimal withdrawals. The contract that we consider is very similar in
  specifications to the real product marketed in Australia, see for example \cite{MLCproduct2014} and \cite{AMPNorthPDS2014}. Numerical difficulties
   encountered in pricing this VA rider are common across other VA guarantees and at the same time comprehensive numerical pricing results for this product are not available in the literature. These results (reported for a range of parameters) can serve as a benchmark for practitioners and researchers developing numerical pricing of VA riders.

In the next section,  a  general specification of VA riders is given.
In Section \ref{stochmodel_sec} we discuss stochastic models
used for pricing these products. Section
\ref{VArider_types_sec} provides precise specification for some
popular VA riders. In Section \ref{PriceStochControl_sec} we outline
the calculation of the fair price and fair fee as a solution of an optimal
stochastic control problem. Section \ref{sec_alg} presents the
numerical methods and algorithms for pricing VA riders.
 In Section \ref{NumericalResults_sec}
we present numerical results  for the fair fees
  of GMAB rider. Concluding remarks are given in Section \ref{conclusion_sec}.

\section{VA rider contract specification}\label{VAcontractspec_sec}
Consider a VA contract with some guarantees for  living and
death benefits purchased by an $x$-year old individual at time
$t_0=0$ with the up-front premium invested in a risky asset (e.g. a
mutual fund), denoted as $S(t)$ at time $t\ge 0$.
The VA rider specification includes dates when events such as withdrawal, ratchet (\emph{step-up}), bonus (\emph{roll-up}), death benefit payment, etc. may occur. Precise definitions of these events depend on the contract and corresponding examples will be provided in Section \ref{VArider_types_sec}.
We
assume that the withdrawal can only take place on the set of the ordered
\emph{event times} $\mathcal{T}=\{t_1,\ldots,t_N\}$, where $T=t_N$  is the contract maturity. Also, the set of policy
anniversaries when the ratchet may occur is denoted as $\mathcal{T}_r$ and is assumed to be a subset of $\mathcal{T}$.
For simplicity on notation we
assume that all other events may only occur on the withdrawal dates.  The value of VA contract with
guarantees at time $t$ is determined by the three main state
variables.
\begin{itemize}
\item \emph{Wealth account} $W(t)$, value of the investment account which is linked to the risky asset $S(t)$ and modelled as stochastic process.
\item \emph{Guarantee account} $A(t)$, also referred in the literature as \emph{benefit base}. It is not changing between event times but can be stochastic via stochasticity in $W(t)$ at the event times depending on the contract features.
\item Discrete state variable $I_n\in \{1,0,-1\}$ corresponding to the states of policyholder is being alive at $t_n$, died during $(t_{n-1},t_n]$, or died before or at $t_{n-1}$ correspondingly. Denote the death probability during $(t_{n-1}, t_n]$ as $q_n=\Pr[I_n=0|I_{n-1}=1]$, i.e. $\Pr[I_n=1|I_{n-1}=1]=1-q_n$. Note that $q_n$ depends on the age of the contract holder at $t_n$ and thus depends on the age $x$ at $t_0=0$.
\end{itemize}

Other state variables are needed if the interest rate and/or volatility are stochastic but these are not affected at the contract event times and typically do not enter formulas for the contract cashflows; these will not be considered explicitly. Extra state variable is also required to track a tax free base to account for taxes; this will be considered in Section \ref{tax_sec}. In principle, different guarantees included in VA may have different benefit base state variables. For notational simplicity and also from practical perspective, we assume that all guarantees in VA are linked to the same benefit base account.

Initially, $W(0)$ and $A(0)$ are set equal to the upfront premium.  The contract holder is allowed to take withdrawal
$\gamma_n$ at time $t_n$, $n=1,\ldots,N-1$.
Denote the values of the benefit base just before and just after $t_n$ as $A(t_{n}^{-})$ and $A(t_{n}^{+})$ respectively, and similarly for the wealth account $W(t_{n}^{-})$ and $W(t_{n}^{+})$.

The contract product specification determines:
\begin{itemize}
\item The contractual (guaranteed) withdrawal amount $G_n$ for the period $(t_{n-1},t_n]$ that may depend on the benefit base $A(t_n^-)$ and/or $W(t_n^-)$.
\item Jump conditions at the event times relating state variables before and after the event, subject to withdrawals $\gamma_n$ belonging to an admissible space $\mathcal{A}_n$:
\begin{eqnarray}
W(t_n^+):&=&h^W_n\left(W(t_n^-),A(t_{n}^{-}),\gamma_n\right),\\
A(t_n^+):&=&h^A_n\left(W(t_n^-),A(t_{n}^{-}),\gamma_n\right),\\
\gamma_n&\in& \mathcal{A}_n\left(W(t_n^-),A(t_{n}^{-})\right),
\end{eqnarray}
where $h^W_n(\cdot)$ and $h^A_n(\cdot)$ are some functions that may also depend on the fee, penalty and annual step-up parameters. For example, if only a \emph{ratchet} is possible at $t_n\in \mathcal{T}_r$ and no other contract events, then
$$A(t_n^+)=\max\left(A(t_{n}^{-}),W(t_{n}^{-})\times 1_{t_n\in \mathcal{T}_r}\right).$$
In practice, several events such as withdrawal, ratchet, bonus, etc. may occur at the same time $t_n$, and the contract specification determines the order of this events.
\item The payout $P_T(W,A)$ at the contract maturity if policyholder is alive at $t=T$.
\item The payout $D_n(W,A)$ to the beneficiary at $t_n$ in the case of the policyholder death during $(t_{n-1},t_n]$, $n=1,\ldots,N$.
\item The cashflow received by the policyholder $\widetilde{f}_n(W(t_n^-),A(t_{n}^{-}),\gamma_n)$ at the event times $t_n$, $n=1,\ldots,N-1$, that might be different from $\gamma_n$ due to penalties.
\end{itemize}

The specification details typically vary across different companies and are difficult to extract from the very long product specification documents. Moreover, results for specific GMxB riders presented in academic literature often refer to different specifications.

Once the above conditions, i.e. functions $h^W_n(\cdot)$, $h^A_n(\cdot)$, $P_T(\cdot)$, $P_D(\cdot)$, $\widetilde{f}_n(\gamma_n)$  and admissible range for withdrawal $\mathcal{A}_n$ are specified by the contract design, and a specific stochastic evolution of the state variables is assumed within $(t_{n-1},t_{n})$, $n=1,\ldots,N$, then pricing of the contract can be accomplished by numerical methods. In particular, if withdrawals are optimal then pricing can be accomplished by PDE, direct integration or regression based MC methods. If withdrawals are deterministic, then standard MC along with PDE and direct integration methods can be used. The use of a particular numerical technique is determined by the complexity of the underlying stochastic model.

\section{Stochastic Model}\label{stochmodel_sec}
Commonly in the literature, stochastic models for the financial risky asset $S(t)$ underlying the VA rider assume that there is no arbitrage in the
financial market which means that there is a risk-neutral measure
$\mathbb{Q}$ under which payment streams can be valued as expected
discounted values. Moreover, this means that the cost of portfolio replicating
the contract is given by its expected discounted value under
$\mathbb{Q}$. Hence, the fair price of the contract can be expressed as an
expectation of the contract discounted cashflows with respect to
$\mathbb{Q}$. Some models considered in the literature assume that
the financial market is complete which means that the risk-neutral
measure $\mathbb{Q}$  is unique. It is also assumed that market has a risk-free asset that accumulates continuously at risk free interest rate.
These are typical assumptions in
the academic research literature on pricing financial derivatives, for a good textbook in this area we refer the reader to e.g. \cite{bjork2004arbitrage}.

Regarding the mortality risk, it is assumed that it is fully diversified via selling the contract to many policyholders. In the case of systemic (undiversified) mortality risk, the risk-neutral fair value can be adjusted using an actuarial premium principle, see e.g. \cite{GaillardetzLakhmiri2011}. Another common assumption is that mortality and financial risks are independent.

A benchmark model commonly considered in the literature on pricing VA riders is the well-known Black-Scholes
dynamics for the reference portfolio of assets $S(t)$ that under the risk-neutral measure $\mathbb{Q}$ is known to be
\begin{equation}\label{underlyingasset_process_eq}
dS(t)=r(t) S(t) dt+\sigma(t) S(t) dB(t).
\end{equation}
Here, $B(t)$ is the standard Wiener process, $r(t)$ is the risk free
interest rate and $\sigma(t)$ is the volatility. Under this model
the financial market is complete. Without loss of generality, the model parameters can be
assumed to be piecewise constant functions of time for time
discretization $0=t_0<t_1<\cdots<t_N=T$. Denote corresponding asset
values as $S(t_0),\ldots,S(t_N)$ and risk free interest rate and
volatility as $r_1,\ldots,r_N$ and $\sigma_1,\ldots,\sigma_N$
respectively. That is, $\sigma_1$ is the volatility for $(t_0,t_1]$;
$\sigma_2$ is the volatility for $(t_1,t_2]$, etc. and similarly for
the interest rate.

 Pricing VA riders in the case of extensions of the above model to the stochastic interest rate and/or stochastic volatility have been developed in e.g. \cite{ForsythVetzal2014}, \cite{LuoShevchenko2016GMWBstInterest}, \cite{bacinello2011unifying}, \cite{huang2015}.

Regarding mortality modelling, the standard way is to use
official Life Tables to estimate the death probability
$q_n=\Pr[I_n=0|I_{n-1}=1]$ during $(t_{n-1}, t_n]$. Life Tables
provide annual death probabilities for each age and gender in a
given country; probabilities for time periods within a year can be
found by e.g. linear interpolation, see
\cite{LuoShevchenkoGMWDB2015}. Instead of a Life Table, stochastic
mortality models such as the  benchmark Lee-Carter model introduced
in \cite{Lee-Carter} can also be used to forecasts the required death
probabilities (accounting for
systematic mortality risk).



 For a given process of risky asset $S(t)$, $t\ge 0$, the value of the wealth account $W(t)$ evolves as
\begin{equation}\label{wealth_process_eq}
\begin{split}
W(t_n^-)&=\frac{W(t_{n-1}^+)}{S(t_{n-1})}S(t_n) e^{-\alpha dt_n},\\
W(t_n^+)&=\max(W(t_n^-)-\gamma_n,0),\;\; n=1,2,\ldots,N,
\end{split}
\end{equation}
where $dt_n=t_n-t_{n-1}$ and $\alpha$ is the annual fee continuously
charged by contract issuer for the provided guarantee. In the case of $S(t)$
following the geometric Brownian motion process
(\ref{underlyingasset_process_eq}), we have
$$ S(t_n)=S(t_{n-1})e^{(r_n-\frac{1}{2}\sigma^2_n)dt_n+\sigma_n \sqrt{dt_n} z_n},$$
  where $z_1,\ldots,z_N$ are independent and identically distributed standard Normal random variables.

In practice, the guarantee fee is charged discretely and proportional to the wealth account that can easily be incorporated into the wealth process (\ref{wealth_process_eq}). Denoting the discretely charged fee with the annual basis as $\widetilde{\alpha}$, the wealth process becomes
\begin{equation}\label{wealth_process_discrfee1}
\begin{split}
W(t_n^-)&=\frac{W(t_{n-1}^+)}{S(t_{n-1})}S(t_n), \\
W(t_n^+)&=\max\left(W(t_n^-)(1-\widetilde{\alpha} dt_n)-\gamma_n,0\right),\;\; n=1,2,\ldots,N.
\end{split}
\end{equation}
Typically, the difference between continuously and discretely charged
fees is not material as observed in our numerical results given in
Section \ref{NumericalResults_sec}.

Another popular fee structure corresponds to fees charged as a proportion of the benefit base, so that
\begin{equation}\label{wealth_process_discrfee2}
\begin{split}
W(t_n^-)&=\frac{W(t_{n-1}^+)}{S(t_{n-1})}S(t_n),\\
W(t_n^+)&=\max\left(W(t_n^-)-A(t_n^-)\widetilde{\alpha} dt_n -\gamma_n,0\right),\;\; n=1,2,\ldots,N.
\end{split}
\end{equation}
Here, it is assumed that discrete fees are deducted before withdrawal but
it can be vice versa depending on the contract specifications.

For simplicity, we do not consider management fees $\alpha_m$ charged by a mutual fund for managing the investment portfolio.
 If management fees $\alpha_m$ is given exogenously, then it will have an impact on the fair fee $\alpha$ that should by
 charged by the VA guarantee issuer. This can be accomplished as described in e.g. \cite{ForsythVetzal2014} and can be easily
 incorporated in the framework outlined in our paper. Obviously, $\alpha$ will be larger for given $\alpha_m>0$ comparing to the
 case $\alpha_m=0$. The management fees reduce the performance of the investment account thus increasing the value of the guarantee as
  reported in e.g. \cite{ChenVetzalForsyth2008} for GMWB or \cite{ForsythVetzal2014} for  GLWB. They commented that insurers wishing
  to provide the cheapest guarantee could provide the guarantee on the corresponding inexpensive exchange traded index
  fund rather than on a managed  mutual fund account with extra fees.

\section{VA riders}\label{VArider_types_sec}
There are many different specifications for GMWB, GLWB, GMAB, GMIB and GMDB in the industry and academic literature. In this section we provide a mathematical formulation for some standard VA rider setups.
We assume that the guarantee fee $\alpha$ is charged continuously.
If the fee is charged discretely (and before withdrawal and other contract events), then one should make the following adjustment to the formulas in this section:
$$W(t_n^-)\rightarrow W(t_n^-)(1-\widetilde{\alpha} dt_n),$$
if the fee is proportional to the wealth account  and
  $$W(t_n^-)\rightarrow \max(W(t_n^-)-A(t_n^-)\widetilde{\alpha} dt_n,0),$$
  if the fee is proportional to the benefit base.

\subsection{GMWB}
A VA contract with GMWB
promises to return at least the entire initial investment through cash withdrawals during
the policy life plus the remaining account balance at maturity, regardless of the
portfolio performance.
Often in academic literature, the studied GMWB type has a very simple
structure, where the penalty is applied to the cashflow paid to the
contract holder, while the benefit base is reduced by the full
withdrawal amount. Specifically,
\begin{equation}\label{guarantee0_eq}
A(t_n^+):=h_n^A(W(t_{n}^{-}),A(t_{n}^{-}),\gamma_n)=A(t_{n}^{-})-\gamma_n,
\end{equation}
with $\gamma_n\in\mathcal{A}_n$, $\mathcal{A}_n=[0,A(t_{n}^{-})]$;  and cashflow paid to the contract holder is
\begin{equation}\label{eqn_penalty0}
\widetilde{f}_n(W(t_n^-),A(t_{n}^{-}),\gamma_n)=\left\{\begin{array}{ll}
                   \gamma_n, & \mbox{if}\; 0 \leq \gamma_n \leq G_n,\\
                   G_n+(1-\beta)(\gamma_n-G_n),& \mbox{if}\; \gamma_n > G_n,
                 \end{array} \right.
\end{equation}
where $\beta\in [0,1]$ is the penalty parameter for excess withdrawal. The contractual amount is defined as $G_n=W(0)(t_n-t_{n-1})/T$ and the maturity condition is
$$P_T(W(t_N^-),A(t_N^-))=\max(W(t_N^-),\widetilde{f}_n(A(t_N^-))).$$
Note that the above specification does not allow early
surrender which can be included via extending the withdrawal space
$\mathcal{A}_n$. Also, there is no death benefit; it is assumed that beneficiary will maintain the contract if the case of policyholder death.
This contract has only basic features facilitating comparison of results from different academic studies, such as
\cite{Forsyth2008}, \cite{dai2008guaranteed}, \cite{LuoShevchenko2015surrender}, \cite{LuoShevchenkoGMWB2015}.

Specifications common in the industry include cases where the
contractual amount $G_n$ is specified to be different from
$G_n=W(0)(t_n-t_{n-1})/T$ and a penalty is applied to both the
withdrawn amount and the benefit base. For example, specifications used in \cite{bauer2015behavior} to compare with the industry products include:
\begin{equation}\begin{array}{l}
\widetilde{f}_n(W(t_n^-),A(t_{n}^{-}),\gamma_n)=\gamma_n-\delta_{excess}-\delta_{penalty},\\
\delta_{excess}=\beta_n^{\mathrm{e}}\max\left(\gamma_n-\min(A(t_n^-),G_n),0\right),\\
\delta_{penalty}=\beta^{\mathrm{g}}_n(\gamma_n-\delta_{excess})\times 1_{x+t_n<59.5},
\end{array}
\end{equation}
where $x$ is the age of the policyholder in years at $t_0=0$, $\beta_n^{\mathrm{e}}$ and $\beta^{\mathrm{g}}_n$ are excess withdrawal and early withdrawal penalty parameters that can change with time, and
$$
\gamma_n\in\mathcal{A}_n,\; \mathcal{A}_n=\left[0,\max\left(W(t_n^-),\min(A(t_n^-),G_n)  \right)\right].
$$
\cite{bauer2015behavior}  also considered several specifications for the benefit base jump conditions.
\begin{itemize}
\item Specification 1:
 \begin{equation}\label{guarantee1_eq}
A(t_n^+)=\left\{\begin{array}{ll}
\max(A(t_{n}^{-})-\gamma_n,0), & \mbox{if }\gamma_n\leq G_n, \\
\max\left(\min\left(A(t_{n}^{-})-\gamma_n , A(t_n^-)\frac{W(t_n^+)}{W(t_n^-)}\right),0\right), & \mbox{if }\gamma_n> G_n.
\end{array}\right.
\end{equation}
\item Specification 2:
\begin{equation}\label{guarantee2_eq}
A(t_n^+)=\left\{\begin{array}{ll}
\max(A(t_{n}^{-})-\gamma_n,0), & \mbox{if }\gamma_n\leq G_n, \\
\max\left(\min\left(A(t_{n}^{-})-\gamma_n , W(t_n^+)\right),0\right), & \mbox{if }\gamma_n> G_n.
\end{array}\right.
\end{equation}
\item Specification 3:
\begin{equation}\label{guarantee3_eq}
A(t_n^+)=\left\{\begin{array}{ll}
\max(A(t_{n}^{-})-\gamma_n,0), & \mbox{if }\gamma_n\leq G_n, \\
\max\left(A(t_{n}^{-})-G_n,0\right) \frac{W(t_n^+)}{\max(W(t_n^-)-G_n,0)}, & \mbox{if }\gamma_n> G_n.
\end{array}\right.
\end{equation}
\end{itemize}

In addition, a ratchet (reset of the benefit base to the wealth
account if the latter is higher) can apply at anniversary
dates. If it occurs before the withdrawal, then in the above
formulas one should make the following adjustment
$$
A(t_{n}^{-})\rightarrow \max(A(t_{n}^{-}),W(t_{n}^{-})),\;\mbox{if}\; t_n\in\mathcal{T}_r.
$$
If the reset is taking place after the withdrawal, then one should
have
$$
A(t_{n}^{+})\rightarrow \max(A(t_{n}^{+}),W(t_{n}^{+})),\;\mbox{if}\; t_n\in\mathcal{T}_r.
$$

\subsection{GLWB}
GLWB is similar to GMWB but provides guaranteed withdrawal for life; upon death the
remaining wealth account value is paid to the beneficiary. The
contractual withdrawal amount $G_n$ is typically based on a fixed
proportion $g$ of the benefit base $A(t)$, i.e. $G_n=g\times
A(t_{n}^-)(t_n-t_{n-1})$. The benefit base can increase via ratchet
(\emph{step-up}) or bonus (\emph{roll-up}) features. Bonus feature
provides an increase of the benefit base if no withdrawal is made on
a withdrawal date. Complete surrender refers to the withdrawal of
the whole policy account. The withdrawal can exceed the contractual
amount and in this case the net amount received by the policyholder
is subject to a penalty. Under the typical specification considered e.g. in \cite{huang2015}, the cashflow received by the
policyholder is

\begin{equation}\label{eqn_GLWB_penalty}
\widetilde{f}_n(W(t_n^-),A(t_{n}^{-}),\gamma_n)=\left\{\begin{array}{ll}
                   \gamma_n, & \mbox{if}\; 0 \leq \gamma_n \leq G_n,\\
                   G_n+(1-\beta)(\gamma_n-G_n),& \mbox{if}\; \gamma_n > G_n,
                 \end{array} \right.
\end{equation}
$$
\gamma_n\in \mathcal{A}_n,\quad \mathcal{A}_n=[0,\max(W(t_n^-),G_n)],
$$
 where $\beta$ is the penalty parameter for excess withdrawal. The benefit base jump condition, including ratchets and bonus features, is given by
\begin{eqnarray}
A(t_n^+)&=&\max\left(A(t_n^-)(1+b_n),W(t_n^-)1_{t_n\in \mathcal{T}_r}\right)\times 1_{\gamma_n=0}\nonumber \\
&&+\max\left(A(t_n^-), \max(W(t_n^-)-\gamma_n,0)\times 1_{t_n\in \mathcal{T}_r}\right)\times 1_{0<\gamma_n\le G_n}\nonumber\\
&&+\max\left( A(t_n^-)\frac{W(t_n^-)-\gamma_n}{W(t_n^-)-G_n}, (W(t_n^-) - \gamma_n)\times 1_{t_n\in \mathcal{T}_r} \right)\times 1_{G_n<\gamma_n\le W(t_n^-)},
\end{eqnarray}
where $b_n$ is the bonus rate parameter that may change in time.
Finally, if the policyholder dies during $(t_{n-1},t_n]$, the beneficiary receives a death benefit payment $D_n(W(t_n^-),A(t_n^-))=W(t_n^-)$ and $t_N$ corresponds to the maximum age beyond which survival is deemed
impossible.

\subsection{GMAB}\label{GMAB_spec_sec}
GMAB rider provides a certainty of capital till some maturity
(e.g. 10 or 20 years) and the potential for a capital growth.
Typical GMAB products sold on the market do not impose penalty on
the policyholder withdrawal amount but can penalise the benefit base
(protected capital balance) under some conditions. It is also common to have a ratchet
feature, where
 the protected capital balance increases to the wealth account if the latter is higher on  an anniversary
date. Withdrawals from the account are allowed subject to a penalty. For example, specifications of the product marketed by \cite{MLCproduct2014} and
\cite{AMPNorthPDS2014} in Australia are very close to the following formulation:
\begin{equation}
\widetilde{f}_n(W(t_n^-),A(t_{n}^{-}),\gamma_n)=\gamma_n,
\end{equation}
\begin{equation}\label{guarantee_eq}
A(t_n^+):=h_n^A(W(t_{n}^{-}),A(t_{n}^{-}),\gamma_n)=\left\{\begin{array}{ll}
\max\left(A(t_{n}^{-}),W(t_{n}^{-})\right)-C_n(\gamma_n), & \mbox{if }t_n\in \mathcal{T}_r, \\
  \max\left(A(t_{n}^{-})-C_n(\gamma_n),0\right), & \mbox{otherwise},
\end{array}\right.
\end{equation}
where $C_n(\gamma_n)$ is a penalty function that can be larger than
$\gamma_n$ as defined below,   and  $\gamma_n\in \mathcal{A}_n=[0,W(t_n^-)]$.

The product is offered for the \emph{super}  and \emph{pension} account types. The \emph{super} account
 is designed for an investor being in an accumulation phase, while the \emph{pension} account is for a retired
 investor in an annuitization phase. The difference between the accounts in terms of technical details is only in the penalty
  applied to the protected capital after withdrawals; the super account discourages withdrawals more
 than the pension account. In both cases the penalty is in the form of a reduction
 of the protected capital (benefit base) larger than the withdrawn amount. The penalty
 only applies
 if the wealth account balance is below the protected capital amount.
A super account penalizes any amount of withdrawals, while the
pension account only penalize excessive
 withdrawals.

 Specifically, for a \emph{super account}, the function $C_n(\gamma_n)$ is given by
\begin{equation}\label{eqn_penaltyS}
C_n(\gamma_n)=\left\{\begin{array}{ll}
                   \gamma_n, & \mbox{if}\; W(t_n^-)\ge A(t_n^-), \\
                   A(t_n^-)\gamma_n/W(t_n^-), & \mbox{if}\; W(t_n^-)< A(t_n^-),
                 \end{array} \right.
\end{equation}
and for a \emph{pension account}, the penalty is
\begin{equation}\label{eqn_penaltyP}
C_n(\gamma_n)=\left\{\begin{array}{ll}
                   \gamma_n, & \mbox{if}\; W(t_n^-) \geq A(t_n^-)\;\;\text{or}\;\; \gamma_n \leq G_n,\\
                   A(t_n^-)\gamma_n/W(t_n^-), & \mbox{if}\; W(t_n^-)< A(t_n^-)\;\;\text{and}\;\; \gamma_n > G_n.
                 \end{array} \right.
\end{equation}

That is, the penalty for the pension account applies only if the
 wealth account balance is below the protected capital amount {\it and} the withdrawal is above a pre-determined amount $G_n$.

Finally, the terminal condition is given by
$$P_T(W(t_N^-),A(t_N^-))=\max(W(t_N^-),A(t_N^-)).$$

 A total withdrawal of the wealth account balance effectively terminates the contract, as the penalty mechanism ensures
  the protected capital is always exhausted to zero by a complete withdrawal.

\subsection{GMIB}
At maturity, the holder of GMIB rider can select to take a lump sum of the wealth account $W(T)$ or annuitise this amount
at annuitization rate $\ddot{a}_T$ current at maturity or annuitize the benefit base $A(T^-)$ at pre-specified annuitization
rate $\ddot{a}_g$. Annuitization rate is defined as the price of an annuity paying one dollar each year. If the account value is below the
 benefit base, then the customer cannot take $A(T^-)$ as a lump sum but only as an annuity at pre-specified rate. Thus, the payoff of VA with GMIB at time $T$ is

$$
P_T(W(t_N^-),A(t_N^-))=\max\left(W(t_N^-), A(t_N^-)\frac{\ddot{a}_T}{\ddot{a}_g}\right).
$$
The benefit base may include roll-ups and ratchets.
Again, this rider can be offered jointly with other riders. For example, it can be part of GMWB or GMAB contract maturity conditions.
For discussion and pricing of GMIB in academic  literature,  see \cite{marshall2010valuation} and \cite{bauer2008universal}.

\subsection{GMDB}
GMDB rider provides a death benefit if the policy holder death occurs before or at the contract maturity. Assuming that if the policyholder dies during $(t_{n-1},t_n]$, then the beneficiary will be paid an amount $D_n(\cdot)$ at $t_n$, where some of the common death benefit types are:
\begin{eqnarray}
D_n(W(t_n^-),A(t_n^-))=\left\{\begin{array}{ll}
\max(A(t_n^-),W(t_n^-)),& \mbox{death benefit type 0},\\
W(0),& \mbox{death benefit type 1}, \\
\max(W(0),W(t_n^-)), & \mbox{death benefit type 2},\\
W(t_n^-),& \mbox{death benefit type 3}.\\
\end{array}
\right.
\end{eqnarray}
Some providers adjust the initial premium $W(0)$ for inflation in
the death benefit. For some policies, the death benefit type may
change at some age, e.g. death benefit type 1 or type 2 may change
to type 0, effectively making the death benefit expiring at some age
(e.g. at the age of 75 years). The death benefit can be provided
on top of some other guarantees and the contract may provide a
\emph{spousal continuation option} that allows a surviving spouse to
continue the contract. The contract may have accumulation phase where the death benefit may increase, and continuation phase where the death benefit remains constant.

Pricing GMDB has been considered in e.g. \cite{milevsky2001titanic}, \cite{belanger2009valuing}, \cite{LuoShevchenkoGMWDB2015}.

 \section{Fair Pricing}\label{PriceStochControl_sec}
 Denote the state vector at time $t_n$ before the withdrawal as $X_n=(W(t_n^-),A(t_n^-),I_n)$ and $\bm{X}=(X_1,\ldots,X_N)$. Given the withdrawal strategy $\bm{\gamma}=(\gamma_1,\ldots,\gamma_{N-1})$,
the present value of the overall payoff of the VA contract with a guarantee is a function of the state vector
\begin{equation}\label{total_payoff_eq}
H_0(\bm{X},\bm{\gamma})= B_{0,N} H_N(X_N)+\sum_{n=1}^{N-1} B_{0,n} f_n(X_n,\gamma_n).
\end{equation}
Here,
\begin{equation}
H_N(X_N)=P_T\left( W(T^-),A(T^-)\right)\times 1_{I_n=1}+D_N\left(W(T^-),A(T^-)\right)\times 1_{I_n=0}
\end{equation}
is the cashflow at the contract maturity
 and
\begin{equation}
f_n(X_n,\gamma_n)= \widetilde{f}_n(W(t_n^-),A(t_{n}^{-}),\gamma_n)\times 1_{I_n=1}+D_n \left(W(t^-_n),A(t_n^-)\right)\times 1_{I_n=0}
\end{equation}
is the cashflow at time $t_n$.
Also, $B_{i,j}$ is the discounting factor from $t_j$ to $t_i$ \begin{equation}
B_{i,j}=\exp\left(-\int_{t_i}^{t_j}r(t)dt\right),\;t_j>t_i.
\end{equation}

\subsection{Pricing as Stochastic Control Problem}
Let $Q_t(W,A)$ be the price of the VA contract with a guarantee at time $t$, when $W(t)=W$, $A(t)=A$ and policyholder is alive.
For simplicity of notation, if the policyholder is alive, we drop
mortality state variable $I_n=1$ in the function arguments.
Assume that financial risk can be eliminated via continuous hedging. Also assume that mortality risk is fully diversified via selling the contract to  many people of the same age,  i.e.
 the average of the contract payoffs $H_0(\bm{X},\bm{\gamma})$ over $L$ policyholders converges to $\mathrm{E}_{t_0}^{\mathbb{I}}[H_0(\bm{X},\bm{\gamma})]$ as $L\rightarrow \infty$, where $\mathbb{I}$ is the real probability measure corresponding to the mortality process $I_1,I_2,\ldots$. Then
 the contract price under the given withdrawal strategy $\bm{\gamma}$ can be calculated as
\begin{equation}\label{GMWDB_general_eq}
Q_0\left(W(0),A(0)\right)=\mathrm{E}_{t_0}^{\mathbb{Q},\mathbb{I}}\left[H_0(\bm{X},\bm{\gamma})\right].
\end{equation}
Here, $\mathrm{E}_{t}^{{\mathbb{Q}},\mathbb{I}}[\cdot]$ denotes an expectation with respect to the state vector $\bm{X}$, conditional on information available at time $t$, i.e. with respect to the financial risky asset process under the risk-neutral probability measure ${\mathbb{Q}}$ and with respect to the mortality process under the real probability measure $\mathbb{I}$.
Then the fair fee value of $\alpha$ to be charged for VA guarantee corresponds to $Q_0(W(0), A(0))=W(0)$. That is, once a pricing of $Q_0(W(0), A(0))$ for a given $\alpha$ is developed, then a numerical root search algorithm is required to find the fair fee.

The withdrawal strategy $\bm\gamma$ can depend on time and state variables and is assumed to be given when price of the contract is calculated in (\ref{GMWDB_general_eq}).
The withdrawal strategies are classified as \emph{static}, \emph{optimal}, and \emph{suboptimal}.
\begin{itemize}
\item \textbf{Static strategy.} Under this strategy, the policyholder decisions are deterministically determined at the beginning of the contract and do not depend on the evolution of the wealth and benefit base accounts. For example, policyholder withdraws at the contractual rate only.

\item \textbf{Optimal strategy.} Under the optimal withdrawal strategy, the decision on the withdrawal amount $\gamma_n$ depends on the information available at time $t_n$, i.e. depends on the state variable $X_n$. The optimal strategy is calculated as
\begin{equation}\label{optimalstrategy_eq}
\bm{\gamma}^\ast(\bm{X})=\underset{\bm{\gamma\in\mathcal{A}}}{\mathrm{argsup}} \mathrm{E}_{t_0}^{{\mathbb{Q}},\mathbb{I}}\left[H_0(\bm{X},\bm{\gamma})\right],
\end{equation}
where the supremum is taken over all admissible strategies $\bm\gamma$. Any other strategy $\bm{\gamma}(\bm{X})$ different from $\bm{\gamma}^\ast(\bm{X})$ is called \emph{suboptimal}.

\end{itemize}


%
%
%


Given that the state variable $\bm{X}=(X_1,\ldots,X_N)$ is a Markov
process and the contract payoff is represented by the general formula
(\ref{total_payoff_eq}),  calculation of the contract value (\ref{GMWDB_general_eq}) under the
optimal withdrawal strategy (\ref{optimalstrategy_eq}) is a standard
optimal stochastic control problem  for a \emph{controlled Markov
process}. Note that, the control variable $\gamma_n$ affects the transition law of the underlying wealth $W(t)$ process from $t_n^-$ to $t_{n+1}^-$ and thus the process is controlled. For a good textbook treatment of stochastic control
problems in finance, see \cite{bauerle2011markov}. This type of
problems can be solved recursively to find the contract value
$Q_{t_n}(x)$ at $t_n$ when $X_n=x$ for $n=N-1,\ldots,0$ via the backward induction \emph{Bellman equation}
\begin{equation}\label{Bellman_eq}
Q_{t_n}(x)=\sup_{\gamma_n\in
\mathcal{A}_n}\left(f_n(x,\gamma_n)+ B_{n,n+1}\int
Q_{t_{n+1}}(x^\prime)K_{t_n}(dx^\prime|x,\gamma_n) \right),
\end{equation}
starting from the final condition $Q_T(x)=H_N(x)$. Here,
$K_{t_n}(dx^\prime|x,\gamma_n)$ is  the stochastic kernel
representing probability to reach state in $dx^\prime$ at time
$t_{n+1}$ if the withdrawal (\emph{action}) $\gamma_n$ is applied in
the state $x$ at time $t_n$.
Obviously, the above backward induction can also be used to  calculate the fair contract price in the case of static strategy $\bm\gamma$; in this case the space of admissible  strategies $\mathcal{A}_n$ consists only one pre-defined value and $\sup(\cdot)$ becomes redundant.

For clarity, denote ${Q}_{t_n^-}(\cdot)$ and ${Q}_{t_n^+}(\cdot)$
the contract values just before and just after the event time $t_n$  respectively. Then, after
calculating expectation with respect to the mortality state variable
$I_{n+1}$ in (\ref{Bellman_eq}), the required backward recursion can
be rewritten explicitly as
\begin{eqnarray}\label{recursion_eq}
{Q}_{t_n^+}\left(W,A\right)&=&(1-q_{n+1})\mathrm{E}_{t_n^+}^{\mathbb{Q}}\bigg[B_{n,n+1}{Q}_{t_{n+1}^{-}}\left(W(t_{n+1}^{-}),A(t_{n+1}^{-})\right)\left|
W,A\right.\bigg]\nonumber\\
&&+q_{n+1}\mathrm{E}_{t_n^+}^{\mathbb{Q}}\bigg[B_{n,n+1}D_{n+1}\left(W(t_{n+1}^{-}),A(t_{n+1}^{-})\right)|W,A\bigg]
\end{eqnarray}
with the jump condition
\begin{equation}\label{jump_cond_eq}
{Q}_{t_n^{-}}\left(W,A\right)=
\underset{\gamma_{n}\in\mathcal{A}_n}{\max}\left(
\widetilde{f}_n(W,A,\gamma_n)+{Q}_{t_n^+}\left(h_n^W(W,A,\gamma_n),h_n^A(W,A,\gamma_n)\right)\right).
\end{equation}

This recursion is solved
for $n={N}-1,{N}-2,\ldots,0$, starting from the maturity condition $
{Q}_{t_{{N}}^{-}}\left(W,A\right)=P_T(W,A)$.

\subsection{Alternative Solution}
Given that the mortality and financial asset processes are assumed independent, and the withdrawal decision does not affect mortality process, one can calculate the expected value of the payoff (\ref{total_payoff_eq}) with respect to the mortality process, $\widetilde{H}_0(\bm{W},\bm{A})=\mathrm{E}_{t_0}^\mathbb{I}\left[H_0(\bm{X},\bm{\gamma})\right]$, and then calculate the price under the optimal strategy $\sup_{\bm{\gamma}} \mathrm{E}_{t_0}^{\mathbb{Q}}[\widetilde{H}_0(\bm{W},\bm{A})]$ or under the given strategy $\mathrm{E}_{t_0}^{\mathbb{Q}}[\widetilde{H}_0(\bm{W},\bm{A})]$. It is easy to find that
\begin{eqnarray}\label{total_payoff_eq2}
\widetilde{H}_0(\bm{W},\bm{A})&=& B_{0,N}\bigg( p_N P_T\left( W(T^-),A(T^-)\right)+q_Np_{N-1}D_N\left(W(T^-),A(T^-)\right)\bigg)\nonumber\\
&&+\sum_{n=1}^{N-1} B_{0,n}\left( p_n\widetilde{f}_n(W(t_n^-),A(t_{n}^{-}),\gamma_n)+p_{n-1}q_n D_n \left(W(t^-_n),A(t_n^-)\right)\right),
\end{eqnarray}
where $p_n=\Pr[\tau>t_{n}|\tau>t_0]$ and $q_n p_{n-1}=\Pr[t_{n-1}<\tau\le t_n|\tau>t_0]$ for random death time $\tau$, i.e. $p_n=p_{n-1}(1-q_n)$. Note that, previously we defined $q_n=\Pr[t_{n-1}<\tau\le t_n|\tau>t_{n-1}]$.

The payoff (\ref{total_payoff_eq2}) has the same general form as the payoff (\ref{total_payoff_eq}). Thus, the optimal stochastic control problem $\Psi_{t_0}(W(0),A(0)) =\sup_{\bm{\gamma}} \mathrm{E}_{t_0}^{\mathbb{Q}}[\widetilde{H}_0(\bm{W},\bm{A})]$ can be solved using Bellman equation (\ref{Bellman_eq}) leading to the following explicit recursion
\begin{eqnarray}\label{recursion2_eq}
{\Psi}_{t_n^+}\left(W,A\right)&=&\mathrm{E}_{t_n^+}^{\mathbb{Q}}\left[B_{n,n+1}{\Psi}_{t_{n+1}^{-}}\left(W(t_{n+1}^{-}),A(t_{n+1}^{-})\right)\left|
W,A\right.\right],\\
{\Psi}_{t_n^{-}}\left(W,A\right)&=&
\underset{\gamma_{n}\in\mathcal{A}_n}{\max}\bigg(
p_n\widetilde{f}_n(W,A,\gamma_n)+p_{n-1}q_n D_n(W,A)\nonumber\\
&&\quad\quad\quad\quad+{\Psi}_{t_n^+}\left(h_n^W(W,A,\gamma_n),h_n^A(W,A,\gamma_n)\right)\bigg),
\end{eqnarray}
for $n={N}-1,{N}-2,\ldots,0$, starting from $
{\Psi}_{t_{{N}}^{-}}\left(W,A\right)=p_N P_T(W,A)+p_{N-1}q_N D_N(W,A)$.

It is easy to verify that this recursion leads to the same solution $\Psi_{t_0}(W,A)=Q_{t_0}(W,A)$ and the same optimal strategy for $\bm\gamma$ as obtained from the recursion (\ref{recursion_eq}--\ref{jump_cond_eq}), noting that
$\Psi_{t_n^-}(W,A)=p_n Q_{t_n^-}(W,A)+p_{n-1}q_n D_n(W,A)$. The result is somewhat obvious because
\begin{equation}
\sup_{\bm{\gamma}}\mathrm{E}_{t_0}^{{\mathbb{Q}},\mathbb{I}}\left[H_0(\bm{X},\bm{\gamma})\right]=\sup_{\bm{\gamma}}
\mathrm{E}_{t_0}^{{\mathbb{Q}}}\left[\mathrm{E}_{t_0}^{{\mathbb{I}}}\left[H_0(\bm{X},\bm{\gamma})\right]\right].
\end{equation}
Note that, $\sup_{\bm{\gamma}}\mathrm{E}_{t_0}^{{\mathbb{Q}},\mathbb{I}}\left[H_0(\bm{X},\bm{\gamma})\right]\neq \mathrm{E}_{t_0}^{{\mathbb{I}}}\left[\sup_{\bm{\gamma}}
\mathrm{E}_{t_0}^{{\mathbb{Q}}}\left[H_0(\bm{X},\bm{\gamma})\right]\right]$. That is, one cannot find the price under the optimal strategy  conditional on the death time and then average over random death times, that would lead to the result larger than $Q_{t_0}(W,A)$, see \cite{LuoShevchenkoGMWDB2015}.
\subsection{Remarks on Withdrawal Strategy}\label{remark_withdrawalstrategy_sec}
The guarantee fare fee based on the optimal policyholder withdrawal is the worst
case scenario for the issuer, i.e. if the guarantee is hedged then
this  fee will  \emph{ensure} no losses for the issuer (in other
words full protection against policyholder strategy and market
uncertainty). Of course this is under the given assumptions about
stochastic model for the underlying risky asset. If the issuer hedges
continuously but investors deviate from the optimal strategy, then
the issuer will receive a guaranteed profit.

Any strategy different from the optimal is sup-optimal and will
lead to smaller fair  fees. Of course the strategy optimal  in this
sense is not related to the policyholder circumstances. The
policyholder may act optimally with respect to his preferences and
circumstances but it may be different from the optimal strategy
calculated in (\ref{jump_cond_eq}). On the other hand, as noted in
\cite{hilpert2014effect}, secondary markets for equity linked
insurance products (where the policyholder can sell their contracts)
are growing. Thus, financial third parties can potentially generate
guaranteed
 profit through hedging strategies from financial products such as VA riders which are not priced according to the worst case assumption of the
 optimal withdrawal strategy. Thus the development of secondary markets for VA riders would lead to an increase in the fees charged by
  the issuing companies. \cite{knoller2015propensity} undertakes an empirical study  of policyholders behavior in Japanese VA market and they show that the moneyness of the guarantee has the largest explanatory power for the surrender rates.

One way to introduce a reasonable suboptimal withdrawal model is to assume that the policyholder follows a default strategy
withdrawing a contractual amount $G_n$ at each event time $t_n$ unless the extra value from optimal withdrawal is greater than $\theta\times G_n$,
$\theta\ge 0$. Setting $\theta=0$ corresponds to the optimal strategy, while $\theta\gg 1$ leads to the strategy of withdrawals at the contract rate.
 This is the approach considered e.g. in \cite{ForsythVetzal2014} and \cite{ChenVetzalForsyth2008}. More complicated approach would specify a life-cycle utility
 model to determine the strategy optimal for the policyholder with respect to his circumstances and preferences, this is the
 approach studied in \cite{moenig2012optimal,horneff2015optimal, gao2012optimal,steinorth2015valuing}.
In any case, once the strategy is specified (estimated empirically
or by another model), one can use (\ref{jump_cond_eq}) to
 calculate the fair price and fair fee with the admissible strategy space $\mathcal{A}_n$ restricted to the specified strategy.

\subsection{Tax consideration}\label{tax_sec}
Withdrawals from the VA type contracts may attract
country and individual specific government taxes.
\cite{bauer2015behavior} demonstrated that including taxes
significantly affects the value of VA withdrawal guarantees. They developed a \emph{subjective} risk-neutral valuation
methodology and produced results in line with empirical market
prices. Following closely to \cite{bauer2015behavior}, we
introduce an extra state variable $R(t)$ to present the \emph{tax
base} which is the amount that may still be drawn tax-free, and
assume that all event times $t_n\in\mathcal{T}$ are the policy anniversary
dates. The
initial premium is assumed to be post-tax and taxes are applied to
future investment gains (not the initial investment).

Denote a
marginal income tax rate as $\widetilde{\kappa}$ and marginal
capital gain tax from investment outside of VA
contract as $\kappa$. It is assumed that earnings from VA are
treated as ordinary income and withdrawals are taxed on a last-in
first-out basis. Thus if the wealth account $W(t_n^-)$ exceeds the
tax base $R(t_n^-)$, any withdrawal up to $W(t_n^-)-R(t_n^-)$ will
be taxed at the  rate $\widetilde{\kappa}$ and will not affect the tax
base; larger withdraws will not be subject to tax but will reduce
the tax base. Specifically, the tax base will be changed at withdrawal time $t_n$ as
$$
R(t_{n}^+)=R(t_n^-)-\max\left(\gamma_t-\max(W(t_n^-)-R(t_n^-),0),0\right).
$$
The cashflow received by the policyholder will be reduced by taxes
$$
{tax}=\widetilde{\kappa}\min\left(\widetilde{f}_n(W(t_n^-),A(t_{n}^{-}),\gamma_n),\max(W(t_n^-)-R(t_n^-),0)\right),
$$
i.e. one has to make the following change in the contract specifications listed in Section \ref{VArider_types_sec}
$$
\widetilde{f}_n(W(t_n^-),A(t_{n}^{-}),\gamma_n)\rightarrow \widetilde{f}_n(W(t_n^-),A(t_{n}^{-}),\gamma_n)-{tax}.
$$

Using arguments for replicating pre-tax cashlows at $t_n$ with
post-tax cashflows at $t_{n+1}$, it was shown in
\cite{bauer2015behavior} that ${Q}_{t_n^+}(W,A,R)$ should be found
not as the direct expectation (\ref{recursion_eq}) but should be
found as the solution of the following nonlinear equation
\begin{eqnarray}
&&{Q}_{t_n^+}(W,A,R)=\mathrm{E}_{t_n^+}^{\mathbb{Q}}\left[V(t_{n+1}^-)|W,A,R\right]\nonumber\\
&&\quad\quad+\frac{\kappa}{1-\kappa}\mathrm{E}_{t_n^+}^{\mathbb{Q}}
\bigg[\max\left[V(t_{n+1}^-)-B_{n,n+1}{Q}_{t_n^+}(W,A,R),0\right]\bigg|W,A,R \bigg],
\end{eqnarray}
where
\begin{eqnarray}\label{nonlinear_tax_eq}
{V}(t_{n+1}^-)&=&(1-q_{n+1})B_{n,n+1}{Q}_{t_{n+1}^{-}}\left(W(t_{n+1}^{-}),A(t_{n+1}^{-}),R(t_{n+1}^-)\right)\nonumber\\
&&+q_{n+1} B_{n,n+1}D_{n+1}\left(W(t_{n+1}^{-}),A(t_{n+1}^{-}),R(t_{n+1}^-)\right).
\end{eqnarray}
This is referred to as \emph{subjective} valuation from the
policyholder perspective and depends on the investor current
position (including possible offset tax responsibilities) and tax
rates. Numerical examples in \cite{bauer2015behavior} show that the
VA guarantee prices accounting for taxes in the above way are lower than ignoring
the taxes (not surprisingly, because it lead to the suboptimal strategy), making the prices overall more aligned with those
observed in the market.

\section{Numerical Valuation of VA riders}\label{sec_alg}
In the case of realistic VA riders with discrete events such as
ratchets and optimal withdrawals, there are no  closed
form solutions and fair price has to calculated numerically, even in the case of simple geometric Brownian motion process for the risky asset. In general, one
can use PDE, direct integration or regression type MC
methods, where the backward recursion
(\ref{recursion_eq}--\ref{jump_cond_eq}) is solved numerically. Of
course, if the withdrawal strategy is known, then one can
always use standard MC to simulate state variables forward in time
till the contract maturity or policyholder death and average the payoff
cashflows over many independent realizations. This  standard
procedure is well known  and no further discussion is needed.

In this section, we give a brief review of different numerical methods that can be used for valuation of VA riders. Then, we provide detailed description of the direct integration method that can be very efficient and simple to implement, when the transition density of the underlying asset or it's moments between the event times are known in closed form. Finally, in Section \ref{hedging_sec} we present calculation of hedging parameters (referred in the literature as \emph{Greeks}).

\subsection{Numerical algorithms}
Simulation based Least-Squares MC method introduced in \cite{Longstaff2001} is designed for uncontrolled Markov process problems and can
be used to account for the contract early surrender, as e.g. in \cite{bacinello2011unifying}. However, it cannot be used if an optimal withdrawal strategy is involved.
This is because dynamic withdrawals affect
  the paths of the underlying wealth account and one cannot carry out a forward simulation step required for the subsequent regression in the backward induction.
  However, it should be possible to apply control
randomization methods extending Least-Squares MC to handle
the optimal stochastic control problems with controlled Markov
processes, as was  recently developed in \citet{Kharroubi2014}. The
idea is to first simulate the control (withdrawals) and the state
variables forward in time, where the control is simulated
independently from other variables. Then, use regression on the
simulated state variables and control to estimate expected value
(\ref{recursion_eq}) and find the optimal withdrawal using
(\ref{jump_cond_eq}). However, the accuracy and robustness of this
method for pricing withdrawal benefit type products have not been studied yet. As
usual, it is expected that the choice of the basis functions for the required regression step will
have significant impact on the performance. We also note that in
some simple cases of the withdrawal strategy admissible space such as bang-bang (no
withdrawal, withdrawal at the contractual rate, or full surrender), it is
possible to develop other modifications of Least-Squares MC
such as in \cite{huang2015} for pricing of the GLWB rider.

The expected value (\ref{recursion_eq}) can also be calculated using
PDE or direct integration methods. In both cases, the modeller
discretizes the space of the state variables and then calculates the
contract value for each grid point. The PDE for calculation of expected value (\ref{recursion_eq}) under the assumed  risk-neutral process for the risky asset $S(t)$ is easily derived using Feynman-Kac theorem; for a good textbook treatment of this topic, see e.g. \cite{bjork2004arbitrage}. However, the obtained PDE can be difficult or even not practical to solve in the high-dimensional case.
In particular, in the case of geometric Brownian motion process for the risky asset (\ref{underlyingasset_process_eq}), the governing PDE in the period between
the  event times is the one-dimensional
Black-Scholes PDE, with jump conditions at each event time
to link the prices at the adjacent periods. Since the benefit base state variable $A(t)$ remains unchanged within
the interval $(t_{i-1},\;t_i), \;i=1,2,\ldots,N$, the contract value
$Q_t(W,A)$ satisfies the following PDE with no explicit
dependence on $A$,

\begin{equation}\label{eqn_pde}
\frac{\partial Q}{\partial
t}+\frac{\sigma^2}{2}W^2\frac{\partial^2 Q}{\partial
W^2}+(r-\alpha)W\frac{\partial Q}{\partial W}-rQ=0.
\end{equation}

This PDE can be solved numerically using e.g. Crank-Nicholson finite difference scheme for each $A$ backward in time with the jump condition (\ref{jump_cond_eq}) applied at the contract event times.
This has been done e.g. in \cite{dai2008guaranteed} and \cite{Forsyth2008} for pricing GMWB with discrete optimal withdrawals. Of course, if the volatility or/and interest rate are stochastic, then extra dimensions will be added to the PDE making it more difficult to solve. \cite{ForsythVetzal2014} used PDE approach to calculate VA rider prices in the case of stochastic regime-switching volatility and interest rate.

Under the direct integration approach, the expected value
(\ref{recursion_eq}) is calculated as an integral approximated by
summation over the space grid points, see e.g. \cite{bauer2008universal}. More
efficient quadrature methods (requiring less points to approximate
the integral) exist. In particular, in the case of a geometric
Brownian motion process for the risky asset, it is very efficient to
use the Gauss-Hermite quadrature as developed in \cite{LuoShevchenkoGHQC2014} and applied for GMWB pricing in \cite{LuoShevchenkoGMWB2015}. Section \ref{GHQCalgo_sec} provides detailed description of the  method for pricing VA riders in general. This method can be applied when the transition density of the underlying
asset between the event times \emph{or it's moments} are known in
closed form. It is  relatively easy to implement
and computationally faster than PDE method because the latter requires many time steps between the event times.
In \cite{LuoShevchenko2016GMWBstInterest}, this method was also used to calculate GMWB in the case of stochastic interest rate under the Vasicek model.

 In both PDE and direct
integration approaches, one needs some interpolation scheme to
implement the jump condition (\ref{jump_cond_eq}), because state
variables located at the grid points of discretized space do not
appear on the grid points after the jump event. This will be discussed in detail in Section \ref{jumpcond_sec}.
Of course, if the underlying stochastic process is more complicated
than geometric Brownian motion (\ref{underlyingasset_process_eq}) and does not allow efficient
calculation of the transition density or its moments, one can always
resort to PDE method.

In our numerical examples of GMAB pricing in Section \ref{NumericalResults_sec}, we
adapt a direct intergation method based on the  Gauss-Hermite integration quadrature applied on a cubic spline
interpolation, \emph{hereafter referred to as GHQC}. For testing purposes,  we also implemented
Crank-Nicholson finite difference (FD) scheme solving PDE
(\ref{eqn_pde}) with the jump condition (\ref{jump_cond_eq}).



\subsection{Overall algorithm description}
Both PDE and direct integration numerical schemes start from a final condition for the contract value at $t=T^-$. Then, a backward time stepping using
(\ref{recursion_eq}) or solving corresponding PDE gives solution for the contract value at $t=t_{N-1}^+$. Application of the
jump condition (\ref{jump_cond_eq}) to the solution at
$t=t_{N-1}^+$ gives the solution at $t=t_{N-1}^-$ from which
further backward in time recursion gives solution at $t_0$. For simplicity assume that there are only $W(t)$ and $A(t)$ state variables. The numerical algorithm then takes the following key steps.

~

\begin{algorithm_new}[Direct Integration or PDE method]\label{algorithm1}

 ~

\begin{itemize}
\item Step 1. Generate an auxiliary finite grid  $0 = A_1 < A_2 <
\cdots < A_J$ to track the benefit base balance $A$.
\item Step 2.  Discretize wealth account balance $W$ space as $W_0<W_1< \cdots<W_M$ to generate the grid for computing the expectation (\ref{recursion_eq}).
\item Step 3. At $t=t_N$, apply the final condition at each node point $(W_m, A_j)$, $j=1,2,\ldots, J$, $m=1,2,\ldots, M$
to get $Q_{t_N^-}(W,A)$.
\item Step 4. Evaluate integration (\ref{recursion_eq}) for each $A_j$, $j=0,\ldots,J$,  to obtain $Q_{t_{N-1}^+}(W, A)$ either using direct integration or solving PDE. In the case of direct integration method, this involves one-dimensional interpolation in $W$ space to find values of $Q_{t_N^-}(W,A)$ at the guadrature points different from the grid points.
\item Step 5. Apply the jump condition (\ref{jump_cond_eq}) to obtain $Q_{t_{N-1}^-}(W, A)$ for all possible
values of $\gamma_{N-1}$ and find $\gamma_{N-1}$ that maximizes $Q_{t_{N-1}^-}(W,
A)$. In general, this involves a two-dimensional interpolation in $(W,A)$ space.
\item Step 6. Repeat Step 4 and 5 for $t=t_{N-2}, t_{N-3}, \ldots, t_1$.
\item Step 7. Evaluate integration (\ref{recursion_eq}) for the backward time step from $t_1$ to $t_0$  to obtain solution  $Q_0(W,A)$ at $W=W(0)$ and $A(0)$, or may be at several points if these are needed for calculation of some hedging sensitivities such as \emph{Delta} and \emph{Gamma} discussed in Section \ref{hedging_sec}.
\end{itemize}
\end{algorithm_new}

 In our implementation of the direct integration method based on the Gauss-Hermite quadrature for numerical examples in Section \ref{NumericalResults_sec}, we use a one-dimensional cubic spline interpolation required to handle integration in Step 4 and bi-cubic spline interpolation to handle jump condition Step 5.

 If the model has other stochastic state variables (similar to $W$) changing stochastically between the contract event times, such as stochastic volatility and/or stochastic interest rate, then grids for these extra dimensions should be generated and the required integration or PDE to evaluate (\ref{recursion_eq}) will have extra dimensions. Also, extra auxiliary state variables (similar to $A$) unchanged between the contract event times, such as tax base and/or extra benefit base, will require extra dimensions in the grid and interpolation for the jump condition at the event times.

We have to consider the possibility of $W(t)$ goes to zero due to withdrawal
and market movement, thus one has to use the lower bound $W_{0} = 0$. The upper
bound $W_M$ should be set sufficiently far from the initial wealth at time zero
$W(0)$. A good choice of such a boundary could be based on the high quantiles of distribution of $S(T)$. For example, in the case of geometric Brownian motion process (\ref{wealth_process_eq}), one can set conservatively
$$W_{M} = W(0)
e^{|\mathrm{mean}(\ln(S(T)/S(0)))|+5\times\mathrm{stdev}(\ln(S(T)/S(0)))}.$$

Often, it is more efficient to use equally spaced grid in $\ln W$ space. In this case, $W_0$ cannot be set to zero and instead should be set to a very small value (e.g. $W_0=10^{-10})$. Also, for some VA riders, using equally spaced grid in $\ln A$ space is also more efficient.

\subsection{Direct integration method}\label{GHQCalgo_sec}
To compute $Q_0\left(W(0),A(0)\right)$, we have to evaluate the
expectations in the recursion (\ref{recursion_eq}).
Assuming the conditional probability
 density of $W(t_n^-)$ given $W(t_{n-1}^+)$  is known in closed form
$\widetilde{p}_n(w|W(t_{n-1}^+))$, the
required expectation (\ref{recursion_eq}) can be calculated as

\begin{equation}\label{eq_intS}
{Q}_{t_{n-1}^+}\left(W(t_{n-1}^+),
A\right)=\int_0^{+\infty} \widetilde{p}_n(w|W(t_{n-1}^+))
\widetilde{Q}_{t_n^-}(w,A) dw,
\end{equation}
where
$$
\widetilde{Q}_{t_n^-}(w,A)=B_{n-1,n}\left((1-q_n){Q}_{t_n^-}(w,A)+q_n D_n(w,A) \right).
$$

The above integral can be estimated using various numerical integration (\emph{quadrature}) methods.
Note that, one can always find $W(t_n^-)$ as a transformation of the standard normal random variable $Z$ as $$W(t_n^-)=\psi(Z):=F_n^{-1}(\Phi(Z)),$$
where $\Phi(\cdot)$ is the standard normal distribution, and $F_n(\cdot)$ and $F_n^{-1}(\cdot)$ are the distribution and its inverse of $W(t_n^-)$. Then, the integral (\ref{eq_intS}) can be rewritten as

\begin{equation}\label{eq_intSN}
{Q}_{t_{n-1}^+}\left(W(t_{n-1}^+),
A\right)=\frac{1}{\sqrt{2\pi}}\int_{-\infty}^{+\infty} e^{-\frac{1}{2}z^2}\widetilde{Q}_{t_n^-}(\psi_n(z),A) dz.
\end{equation}

This type of integrand is very well suited for the Gauss-Hermite quadrature that for an arbitrary function $f(x)$ gives the following approximation
\begin{equation}\label{eq_GHQ}
\int_{-\infty}^{+\infty}e^{-x^2}f(x)dx \approx \sum_{i=1}^q
\lambda_i^{(q)} f(\xi_i^{(q)}).
\end{equation}
Here, $q$ is the order of the Hermite polynomial, $\xi_i^{(q)}, i =
1,2,\ldots,q$ are the roots of the Hermite polynomial $H_q(x)$, and
the associated weights $ \lambda_i^{(q)}$  are given by
$$\lambda_i^{(q)}= \frac {2^{q-1} q! \sqrt{\pi}} {q^2[H_{q-1}(\xi_i^{(q)})]^2}.$$
This approximate integration works very well if function $f(x)$ is
without singularities and it calculates the integral exactly if
$f(x)$ is represented by a polynomial of degree $2q-1$ or less.

Note that $\widetilde{Q}_t(w,\cdot)$ is known only at the grid points $W_m$,
$m=0,1,\ldots,M$ and interpolation is required to estimate $\widetilde{Q}_t(w,\cdot)$ at the quadrature points. From our experience with pricing different VA guarantees, we recommend the use of the natural cubic spline interpolation
 which is smooth in the first derivative and continuous in the second derivative; and the second derivative is assumed zero for the extrapolation region
 above the upper bound.

Of course it can be difficult to find the distribution $F_n(\cdot)$ and its inverse $F_n^{-1}(\cdot)$ in general.
In the case of geometric Brownian motion process
(\ref{wealth_process_eq}), the transition density $\widetilde{p}_n(\cdot|\cdot)$ is just a lognormal density and
$$W(t_n^-)=\psi_n(Z):=W(t_{n-1}^+)\exp\left((r_n-\alpha-\frac{1}{2}\sigma^2_n
)dt_n+\sigma_n \sqrt{dt_n} Z\right).$$
Then, a straightforward application of the Gauss-Hermite quadrature for the evaluation of integral (\ref{eq_intS}) gives
 \begin{equation}\label{eq_qX}
{Q}_{t_{n-1}^+}\left(W(t_{n-1}^+), A\right) \approx  \frac{
1}{\sqrt{\pi}} \sum_{i=1}^q \lambda_i^{(q)}
\widetilde{Q}_{t_n^-}\left(\psi_n(\sqrt{2}\xi_i^{(q)}),A\right),
\end{equation}
that should be calculated for each grid point $W(t_{n-1}^+)=W_m$, $m=0,1,\ldots,M$.
Often, a small number of quadrature points is required to achieve
very good accuracy; in our numerical examples in the next section we
use $q=9$ but very good results are also obtained with $q=5$.

If the transition density function
from $W(t_{n-1}^+)$ to $W(t_n^-)$ is not known in closed form but
one can find its moments, then the integration can also be done
with similar efficiency and accuracy  by method of matching moments as described
in \citet{LuoShevchenkoGHQC2014,LuoShevchenkoGMWB2015}.
The method also works very well in the two-dimensional case, see e.g. \cite{LuoShevchenko2016GMWBstInterest} where it was applied for GMWB pricing in the case of stochastic interest rate.


\subsection{Jump condition application}\label{jumpcond_sec}
Either in PDE or direct integration method, one has to apply the jump condition (\ref{jump_cond_eq}) at the event times to obtain $Q_{t_n^-}(W,A)$.
%
 For the optimal strategy, we chose a value of withdrawal $\gamma_n\in\mathcal{A}_n$
 maximizing the
value $Q_{t_n^-}(W,A)$.

\begin{figure}[!h]
\captionsetup{width=0.95\textwidth}
\begin{center}
\includegraphics[scale=0.7]{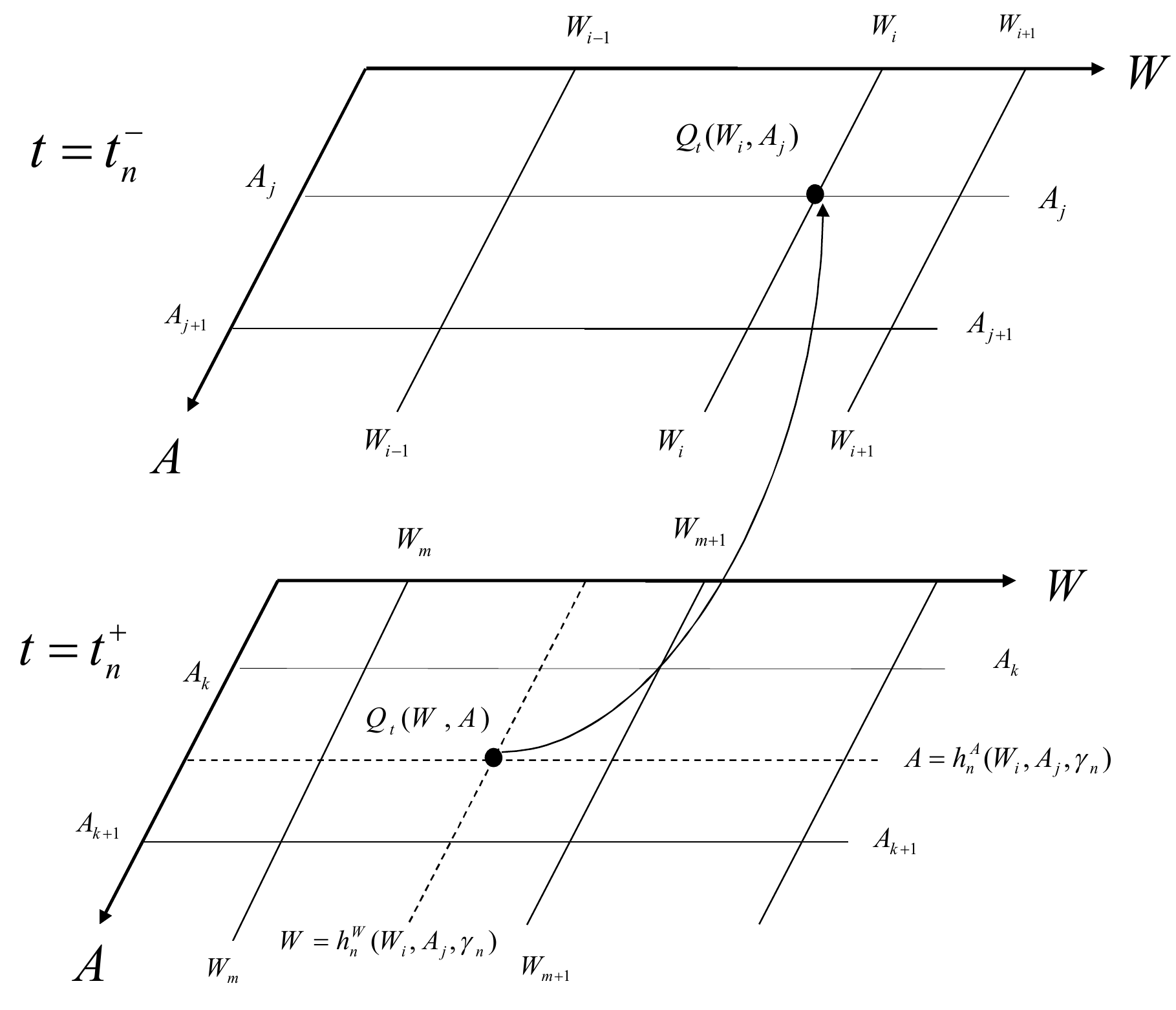}
\caption{{\footnotesize{Illustration of the application of jump
condition. The value $Q_t(W_i,A_j)$ at $t=t_n^-$ and at node point
$(W_i,A_j)$ equals to $Q_t(W,A)$ at $t=t_n^+$ with $W=W_i-\gamma_n$
and  $A=A_j-C(\gamma_n)$. The point $(W,A)$ is located inside the
grid bounded by $(W_m,W_{m+1})$ and $(A_k,A_{k+1})$.
}}}\label{fig_illustration}
\end{center}
\end{figure}

To apply the jump conditions,  an auxiliary finite grid $0 = A_1 <
A_2 < \cdots < A_J = W_{M}$ is used to track the remaining
benefit base state variable $A$. For each $A_j $, we associate a
continuous solution using (\ref{eq_qX}) and interpolation.
In general, as can be seen from (\ref{jump_cond_eq}), the jump
condition makes it impractical, if not impossible, to
ensure the values of $W$ and $A$ after the jump to always fall on a grid
point. Thus a
two-dimensional interpolation is required.
In this work we adopted the bi-cubic spline
interpolation for accuracy and efficiency. Figure
\ref{fig_illustration} illustrates the application of jump
conditions.

 It is natural to form a uniform grid
in $A$ so that optimal withdrawal strategies can be tested on a
constant increment $\delta A=A_{j+1}-A_j$, as has been done
successfully in \cite{LuoShevchenkoGMWB2015} for pricing of a basic GMWB specified by (\ref{guarantee0_eq}--\ref{eqn_penalty0}). However, extensive numerical tests
show that if a uniform grid in $A$ is used for pricing GMAB with ratchets and optimal withdrawals (our numerical example in Section \ref{NumericalResults_sec}), then neither linear
interpolation nor cubic interpolation in $A$ can achieve an
efficient convergence in pricing results. A very fine mesh has to
be used before we see a stable solution, which can take up to
several hours to obtain a fair fee, in sharp contrast to basic GMWB where
less than one minute computer time is required. On the other hand,
if we make the grid in $A$ uniform in $Y=\ln{A}$ and use linear
or cubic interpolation based on variable $Y$, then we obtain a very good
convergence on a moderately fine grid and the CPU time for a fair
fee is about 30 minutes (a few minutes for a fair price).  The CPU
used for all the calculations in this study is
 Intel(R)  Core(TM) i5-2400 @3.1GHz.

As we have already mentioned, a two-dimensional
interpolation has to be used for applying the jump condition. We suggest to use
either a bi-linear interpolation or a bi-cubic spline interpolation,
e.g. see \cite[section 3.6]{Pres92}, in both cases applied on the log-transformed state variables $X=\ln W$ and $Y=\ln A$. For numerical examples in this paper, we
have adapted the more accurate bi-cubic spline interpolation for all
the numerical results.

For uniform grids, the bi-cubic spline is
about five times as expensive in terms of computing time as the
one-dimensional cubic spline.
Suppose the jump condition requires the value $Q(W,A)$ at the point
$(W,A)$ located inside a grid: $W_i\leq W \leq W_{i+1}$ and $A_j\leq
A \leq A_{j+1}$. Equivalently, the point $X=\ln W$ and $Y=\ln A$ is
inside the grid: $\ln W_i =x_i \leq X \leq x_{i+1}=\ln
W_{i+1}$ and $\ln A_j=y_j \leq Y \leq y_{j+1}=\ln A_{j+1}$.
Because the grid is uniform in both $X$ and $Y$ variables, the second
derivatives $\partial^2 Q/\partial X^2$ and  $\partial^2 Q/\partial
Y^2$ can be accurately approximated by the three-point central
difference, and consequently the one-dimensional cubic spline on a
uniform grid involves only four neighboring grid points for any
single interpolation. For the bi-cubic spline, we can first
obtain $Q(\cdot,\cdot)$ at four points $Q(W,A_{j-1})$, $Q(W,A_{j})$,
$Q(W,A_{j+1})$, $Q(W,A_{j+2})$ by applying the one-dimensional cubic
spline on the dimension $X=\log W$ for each point and then we can use
these four values to obtain $Q(W,A)$ through a one-dimensional cubic
spline in $Y=\log A$. Thus five one-dimensional cubic spline
interpolations are required for a single bi-cubic spline interpolation, which involves
sixteen grid points neighboring $(W,A)$ point.


\subsection{Calculating Greeks for hedging}\label{hedging_sec}
Calculation of the contract price in (\ref{recursion_eq}) under the risk neutral probability measure $\mathbb{Q}$  means that
one can find a portfolio replicating the VA guarantee, i.e. perform
hedging eliminating the financial risk. Finding correct hedging depends
on the underlying stochastic model for the risky asset. The basic hedging is the so-called \emph{delta hedging} eliminating randomness due to stochasticity in the underlying risky asset $S(t)$. Here, we use $S(t)$ as a tradable asset to hedge the exposure of the guarantee to the wealth account $W(t)$. One can construct a portfolio consisting of the money market account and $\Delta_S(t)$ units of $S(t)$, so that $\Delta_S(t) S(t)=\Delta_W(t) W(t)$, where $\Delta_W(t)$ is the number of units of the wealth account referred as $Delta$. Denote the value of the VA guarantee as  $U_t(W,A)$
which is just a difference between the contract value with the
guarantee $Q_t(W,A)$ and the value of the wealth account $W$, i.e.
\begin{equation}
Q_t(W,A)=U_t(W,A)+W.
\end{equation}

Then, under the delta hedging strategy, one has to select
$$
\Delta_W(t)=\frac{\partial U_t(W,A)}{\partial W}\Leftrightarrow \Delta_S(t)=\frac{\partial U_t(W,A)}{\partial S}\frac{W}{S}
$$
for time $t$ between the contract event times. Of course if there are extra stochasticities in the model such as stochastic interest rate and/or stochastic volatility, delta hedging will not eliminate risk completely and hedging with extra assets will be required which is model specific. See e.g. \cite{ForsythVetzal2014}, for constructing hedging in the case of regime switching stochastic volatility and interest rate. A popular active hedging strategy in the case of extra stochastic factors is the minimum variance hedging strategy, where $\Delta_W(t)$ is selected to minimize the variance of portfolio's instantaneous changes, e.g. applied in \cite{huang2015} for hedging GLWM in the case of stochastic volatility model. Practitioners also calculate other sensitivities (partial derivatives) of the contract with respect to the interest rate and volatility (referred to as $Rho$ and $Vega$) and even second partial derivatives such as $Gamma=\partial^2 U_t(W,A)/\partial W^2$ to improve hedging strategies. Here, we refer the reader to the standard textbooks in the area of pricing financial derivatives such as \cite{wilmott2013paul} or \cite{hull2006options}.

Numerical estimation of the contract sensitivities (referred to as \emph{Greeks}) is  more difficult than estimation of the contract price.
A general standard approach to calculate Greeks is to perturb the relevant parameter and re-calculate the price. Then one can use a two-point central difference to estimate the first derivatives and a three-point central difference for the second derivatives.
In general, the finite difference PDE (or direct integration) methods generally produce
superior accuracy in calculating Greeks when compared to Monte Carlo method (at least for low dimensions when finite difference method is practical or direct integration is possible).
For Delta and Gamma, the finite difference method (or direct integration method) yields second order accurate values without re-calculating price
using prices already calculated at the uniform grid points.

More accurate calculation of the main Greeks, $Delta$ and  $Gamma$, can be
achieved using the so-called \emph{likelihood method} as follows. The contract price at $t_0$ is calculated
in the last time step $(t_0, t_1)$ in backward induction as an integral (\ref{eq_intS}). Differentiating (\ref{eq_intS}) with respect to $W(0)=w_0$, Delta can be found as
 \begin{eqnarray}\label{Delta_formula}
 \frac{\partial  {Q}_{t_0^+}(w_0,A)}{\partial w_0}&=&\int \widetilde{Q}_{t_1^-}(w,A)\frac{\partial \widetilde{p}_1(w|w_0)}{\partial w_0}dw \nonumber\\\
& =&\int \widetilde{Q}_{t_1^-}(s,A)\frac{\partial \ln\widetilde{p}_1(w|w_0)}{\partial s_0}\widetilde{p}_1(w|w_0)dw \nonumber\\
&=&
{\mathrm{E}}^{\mathbb{Q}}_{t_{0}^+}\left[\widetilde{Q}_{t_1^-}(W,A)\frac{\partial
\ln\widetilde{p}_1(W|w_0)}{\partial w_0}\bigg|w_0,A\right].
\end{eqnarray}
Thus it  can  be  calculated using the same direct integration method as
used for $Q_{t_0}^+(s_0,A)$ with the factor ${\partial
\ln\widetilde{p}_1(W|w_0)}/{\partial w_0}$ added to the integrand.
Similarly, the required derivative  to calculate $Gamma$ can be  found as
\begin{eqnarray}\label{Gamma_formula}
\frac{\partial^2  Q_{t_0^+}(w_0,A)}{\partial w_0^2}&=&\int \widetilde{Q}_{t_1^-}(w,A)\frac{\partial}{\partial w_0}
\left[\frac{\partial \ln\widetilde{p}_1(w|w_0)}{\partial w_0}\widetilde{p}_1(w|w_0)\right]dw \nonumber\\
&=&\int \widetilde{Q}_{t_1^-}(w,A) \left[\left(\frac{\partial \ln\widetilde{p}_1(w|w_0)}{\partial w_0}\right)^2+ \frac{\partial^2 \ln\widetilde{p}_1(w|w_0)}{\partial w_0^2}\right]\widetilde{p}_1(w|w_0)dw \nonumber\\
&=&
{\mathrm{E}}^{\mathbb{Q}}_{t_{0}^+}\left[\widetilde{Q}_{t_1^-}(W,A)\left[\left(\frac{\partial \ln\widetilde{p}_1(w|w_0)}{\partial w_0}\right)^2+ \frac{\partial^2 \ln\widetilde{p}_1(w|w_0)}{\partial w_0^2}\right]\bigg|w_0,A\right].
\end{eqnarray}
Note, the above integrations for Delta and Gamma are only required
for the $(t_0,t_1)$ time step and for a single grid point $W(0)=w_0$. Here, $t_0$ should be understood as the current contract valuation time rather than time when the contract was initiated.
Equivalently, for PDE approach using finite difference method, one can sometimes
derive the corresponding PDEs for the Greeks and solve these
PDEs for the last time step, see e.g. \cite{Tavella2000}. Similarly for Monte Carlo method, simulations used to calculate the price can be used to calculate $Delta$ and $Gamma$ weighted with extra factors under the expectations in (\ref{Delta_formula}) and (\ref{Gamma_formula}).

It is illustrative to show how to derive hedging portfolio under the
basic Black-Scholes  model. Here, we assume that the underlying risky asset $S(t)$ follows
\begin{equation}
dS(t)=\mu(t)S(t)+\sigma(t)S(t)dB^P(t),
\end{equation}
where $B^P(t)$ is the standard Brownian motion under the physical
(\emph{real}) probability measure, and $\mu(t)$ is the real drift. Then the
wealth account evolution is
$$
dW(t)=(\mu(t)-\alpha)W(t)+\sigma(t)W(t)dB^P(t).
$$
Here, we assume a continuously charged fee proportional to the wealth
account but it is not difficult to deal with the case of discretely
charged fees.

To hedge, the guarantee writer takes a long position in $\Delta_S$ units of $S(t)$, i.e. forms a portfolio
$$
\Pi_t=-U_t(W,A)+\Delta_S \times S
$$
By Ito's lemma, the changes of portfolio within $(t_{n-1},t_n)$, $n=1,\ldots,N$ are
\begin{equation}\label{portfolio_change_eq}
d\Pi_t=-\left(\frac{\partial U_t}{\partial t}+\frac{\partial U_t}{\partial W}dW+\frac{1}{2}\sigma^2 S^2\frac{\partial^2 U_t}{\partial W^2} \right) dt-\Delta_S dS + \alpha W dt,
\end{equation}
where the last term $\alpha W dt$ is the fee amount collected over
$dt$. Setting
\begin{equation}\label{delta_eq}
\Delta_S=\frac{\partial U_t(W,A)}{\partial W}\frac{W}{S}=\left(\frac{\partial Q_t(W,A)}{\partial W}-1\right)\frac{W}{S}
\end{equation}
eliminates all the random terms in (\ref{portfolio_change_eq}),
making the portfolio locally riskless. This means that the portfolio
earns at risk free interest rate $r(t)$, i.e. $d\Pi_t=r \Pi_t dt$,
leading to the PDE
\begin{equation}\label{pde_rider_eq}
\frac{\partial U_t}{\partial t}+(r-\alpha)W\frac{\partial U_t}{\partial W}+\frac{1}{2}\sigma^2 S^2\frac{\partial^2 U_t}{\partial W^2}-r U_t -\alpha W=0.
\end{equation}
Substituting $U_t(W,A)=Q_t(W,A)-W$ in the above gives the PDE for
$Q_t(W,A)$, the total value for the contract with the guarantee,
i.e. the same as (\ref{eqn_pde}). Recalling Feynman-Kac theorem, it is easy to see that the stochastic process for $W$ corresponding to this PDE is the risk-neutral process (\ref{underlyingasset_process_eq}).

\section{Numerical Example: GMAB pricing}\label{NumericalResults_sec}
Numerical solutions for pricing VA riders involve many complicated numerical procedures and features. These are more involved when compared to pricing of most exotic derivatives in financial markets. It is important that these solutions are properly tested and validated. As a numerical example for illustration, using direct integration method (GHQC), we calculate accurate prices of
GMAB with possible annual \emph{ratchets} and allowing optimal withdrawals as specified
in Section \ref{GMAB_spec_sec}. With these features, the GMAB  is
very close to the real product marketed in Australia by e.g.
\cite{MLCproduct2014} and \cite{AMPNorthPDS2014}. We assume geometric Brownian motion model for the risky asset (\ref{underlyingasset_process_eq}). When applicable, we compare results with MC and finite difference PDE methods. Numerical
difficulties encountered  in pricing this GMAB rider are common across
other VA guarantees. Also, comprehensive numerical pricing results for this particular product are not available in the literature.
 These validated results (reported for a range of parameters) can serve as a benchmark for practitioners and researchers developing
  numerical pricing of VAs with guarantees.
We consider four GMAB types:
\begin{itemize}
\item[1.] GMAB with the annual ratchets but no withdrawals. In this case, a standard MC method can be used to compare
with GHQC results -- this is a good validation of the
implemented numerical functions related to the \emph{ratchet} feature,
in addition to validating the numerical integration by Gauss-Hermite quadrature.
\item[2.] GMAB with the annual ratchets and a regular withdrawals of a fixed percentage of the wealth account. In this case a standard MC method can also be used to compare with GHQC results -- this is a good validation of implemented numerical functions related to jump conditions due to both \emph{ratchet} and \emph{withdrawal} features. In addition, in order to test the numerical functions related to the application of penalties, we assume a
pension account where the static withdrawal rate is set above the penalty threshold.
\item[3.] GMAB with the optimal quarterly withdrawals and the annual ratchets for a \emph{super account}, where the penalty (\ref{eqn_penaltyS}) is applied on any withdrawal $\gamma_n$ when $W(t_n^-)<A(t_n^-)$.
\item[4.] GMAB with the optimal quarterly withdrawals and the annual ratchets for a \emph{pension account}, where the penalty (\ref{eqn_penaltyP}) is applied if the withdrawal $\gamma_n$ is above the penalty threshold $G_n$  \emph{and} if
$W(t_n^-)<A(t_n^-)$.

\end{itemize}

As a comparison, results from our PDE finite difference implementation
will also be shown for Case 4, the most complicated example among
the four listed above. In addition, we will also calculate results
 for Case 4 in the case of the guarantee fee charged discretely (quarterly). All reported GHQC results are based on $q=9$ quadrature points. We did not observe any material difference in results if $q$ is increased further. Results based on $q=5$ are also very accurate.


\subsection{GMAB with ratchet and no withdrawal}
Consider a GMAB rider with the annual ratchet and no
withdrawals. In this case a standard MC method can be
used to compare with GHQC results which is a good validation of
implemented  numerical functions related to the \emph{ratchet}
feature. Table \ref{tab_feeR1} compares GHQC and MC results for the fare fee $\alpha$ of
GMAB with the annual ratchet for the interest rate $r$ ranging from $1\%$ to $7\%$ and the volatility
$\sigma=10\%$ and $20\%$.
The maximum relative difference between the two methods is $0.76\%$ at
interest rate $r=5\%$ and $\sigma=10\%$. The maximum absolute difference
between the two methods is one basis point at the lowest interest
rate $r=1\%$ where the fee is the highest. On average, the relative
difference is $0.52\%$ and the absolute difference is 0.5 basis
point, which is 5 cents per year on a one thousand dollar account.
The GHQC results are obtained with the mesh size $M=400$ and $J=200$,
and on the average it takes 22 seconds per price (calculation of  the
fare fee requires iterations over several prices). The MC results are obtained using 20 million simulations and it takes about 62 seconds per price.

\begin{table}[!h]\begin{center}
\captionsetup{width=0.9\textwidth}
{\footnotesize{\begin{tabular*}{0.9\textwidth}{ccccccccc} \toprule
\multirow{2}{*}{Interest rate, $r$} &\mbox{\;\;}& \multicolumn{3}{c}{$\sigma=10\%$ }   & \mbox{\;\;}& \multicolumn{3}{c}{$\sigma=20\%$} \\
& &   GHQC, bp &    MC, bp  &   $\widetilde{\delta}$ & &  GHQC, bp &    MC, bp  & $\widetilde{\delta}$\\
 \midrule
$1\%$ & & 337.2 &338.2 & $0.30\%$& & 998.7 &999.8 & $0.11\%$\\
$2\%$ & & 186.0 &186.8 & $0.43\%$& & 637.1 &637.7 & $0.09\%$\\
$3\%$ & &116.8 &117.3 & $0.43\%$& & 458.0 &458.5 & $0.11\%$\\
$4\%$ & &77.94 &78.31 & $0.47\%$& & 346.9 &347.5 & $0.17\%$\\
$5\%$ & &53.91 &54.32 & $0.76\%$& & 271.1 &271.6 & $0.18\%$\\
$6\%$ & &38.54 &38.77 & $0.60\%$& & 216.3 &216.7 & $0.18\%$\\
$7\%$ & &28.11 &28.30 & $0.68\%$& & 175.1 &175.3 & $0.34\%$\\
\bottomrule
\end{tabular*}
}} \caption{{\footnotesize{Fair fee $\alpha$ in bp (1 bp=0.01\%) as a
function of the interest rate $r$ for the GMAB rider with the annual ratchet and no withdrawal.
The contract maturity is $T=10$ years.
$\widetilde{\delta}$ is the relative difference between
Monte Carlo (MC) and GHQC method results.}}} \label{tab_feeR1}
\end{center}
\end{table}

The agreement between the two methods at $\sigma=20\%$ is also very good. In absolute terms, the maximum difference between the
two methods is 1.1 basis point at the lowest interest rate $r=1\%$.
In relative terms, the maximum difference between the two methods is
$0.34\%$ at the highest interest rate $r=7\%$. On average, the
relative difference is $0.17\%$ which is significantly smaller than
the corresponding case at $\sigma=10\%$.

\subsection{GMAB with ratchet and static withdrawal}
Consider a GMAB rider with the annual ratchet and a
regular  quarterly withdrawals of a fixed percentage of the wealth account. In this case
a standard MC method can also be used to compare with the GHQC results
which is a good validation of implemented numerical functions
related to jump conditions due to both \emph{ratchet} and \emph{withdrawal}
features. Here, we consider a pension type account with the penalty given by (\ref{eqn_penaltyP}). In this case,
regular withdrawals at a pre-determined percentage level are allowed.
In order to test the numerical functions related to the application
of penalty, we also consider the static withdrawal above a
pre-determined threshold level that will attract a penalty. We set
the withdrawal threshold at $15\%$ of the wealth account per annum,
and the withdrawal frequency is quarterly, i.e. the quarterly withdrawal
threshold is $G_n=3.75\%$ of the wealth account.

In the first test we allow a regular quarterly withdrawal of $3.75\%$ of the wealth account balance, i.e.  $\gamma_n=G_n$ and there is
no penalty on the withdrawals. Table \ref{tab_feeS1} compares GHQC and MC results
for the fare fee $\alpha$.  In relative terms, the maximum difference
between the two methods is $0.08\%$ at interest rate $r=5\%$.  On
average, the relative difference is $0.06\%$ and the absolute
difference is 0.3 basis point. The GHQC results are obtained with the mesh
size $M=400$ and $J=400$, and the MC results are obtained with 20 million simulations per price.

\begin{table}[!h]\begin{center}
\captionsetup{width=0.95\textwidth}
{\footnotesize{\begin{tabular*}{0.95\textwidth}{ccccccccc} \toprule
\multirow{2}{*}{Interest rate, $r$} &\mbox{\;\;}& \multicolumn{3}{c}{15\% annual withdrawal}   & \mbox{\;\;}& \multicolumn{3}{c}{16\% annual withdrawal} \\
& &   GHQC, bp &    MC, bp  &   $\widetilde{\delta}$ & &  GHQC, bp &    MC, bp  & $\widetilde{\delta}$\\
 \midrule
$1\%$& & 1084 & 1085& $0.09\%$& & 185.3 &185.3 & $< 0.01\%$\\
$2\%$& & 669.1 &669.5 & $0.06\%$& & 152.9 &152.9 & $<0.01\%$\\
$3\%$& & 464.1 &464.4 & $0.06\%$& & 126.6 &126.6 & $<0.01\%$\\
$4\%$& & 339.0 &339.2 & $0.06\%$& & 105.1 &105.1 & $<0.01\%$\\
$5\%$& & 255.0 &255.2 & $0.08\%$& & 87.54 &87.51 & $0.03\%$\\
$6\%$& & 195.7 &195.7 & $<0.01\%$& & 73.21 &73.14 & $0.1\%$\\
$7\%$& & 152.1 &152.2 & $0.07\%$& & 61.40 &61.36 & $0.07\%$\\
\bottomrule
\end{tabular*}
}} \caption{{\footnotesize{Fair fee $\alpha$ in bp  as a function of
interest rate $r$ for the GMAB with the annual ratchet and static
quarterly withdrawal of $3.75\%$  and $4\%$ of the wealth account ($15\%$ and $16\%$ annually respectively). The penalty
threshold (pension type account) is set at $15\%$ annually. The contract maturity is
 $T=10$ years and volatility is $\sigma =20\%$. $\widetilde{\delta}$ is the
relative difference between Monte Carlo (MC) and GHQC results.}}}
\label{tab_feeS1}
\end{center}\end{table}

Comparing with Table \ref{tab_feeR1}, the fair fee for the static
withdrawal is about $8\%$ higher than the corresponding
no-withdrawal case at the lowest interest rate $r=1\%$, but it is
about $13\%$ lower than the corresponding no-withdrawal case at
$r=7\%$. We have also tested static $2.5\%$ quarterly withdrawals and obtained
the same pattern: at lower interest rate the fair fee  of static
withdrawal (which is also below the penalty threshold) is higher than the
corresponding no-withdrawal case, and at higher interest it is the
opposite. These differences in the fair fees at relatively low and
high interest rates can be broadly interpreted as follows. At lower
interest rates, where the expected capital growth is relatively slow,
it is better  to perform a regular withdrawal at or below the
penalty threshold and take the protected capital at the maturity. However,
 at higher interest rates, where the expected capital growth
rate is also high, it is beneficial not to carry
 out a regular withdrawal and keep the capital to grow.

 The above test also demonstrates that the MC and GHQC methods agree very
well for pricing GMAB with a static withdrawal not exceeding the
penalty threshold. This confirms the accuracy and efficiency of our numerical implementation of the jump condition using a bi-cubic interpolation in GHQC method.

 In the
second test of static withdrawal, we allow a regular quarterly
 withdrawal of $4\%$ ($16\%$ per annum), i.e. the annual withdrawal rate is  slightly
 higher than the penalty threshold of $15\%$ per annum and there is a penalty applied for each withdrawal.
The GHQC and MC results for this test are also presented in Table \ref{tab_feeS1}.
In absolute terms, the maximum difference between the two methods is
only 0.07 basis point at interest rate $r=6\%$, which is less than 1
cent per year for a one thousand dollar account. In relative terms,
the maximum difference between the two methods is $0.1\%$ at
interest rate $r=6\%$. On average the relative difference is only
$0.03\%$, which shows the two methods  also agree very well in the
case of excessive static withdrawals, where the penalty is applied
for each withdrawal. This is a very convincing validation that
our GHQC implementation of all numerical
functions associated with the jump conditions, including the  bi-cubic  spline interpolation, is correct and accurate. The above tests are very close
 to validation of the entire algorithm in the case of optimal
withdrawals. This is because in pricing the optimal
withdrawal case, exactly the same integration and interpolation functions
are used, and the only extra step required is to find the withdrawal rate maximizing the
price.
Nevertheless, for optimal strategy cases we will carry out some further validations by comparing results between GHQC and finite difference PDE methods.

Comparing Table \ref{tab_feeS1} with Table \ref{tab_feeR1} for the
no-withdrawal case, the fair fee for the static withdrawal in excess
of the penalty threshold is dramatically reduced: it is reduced by
about $80\%$ from the corresponding no-withdrawal case at $r=1\%$,
and it is reduced by about $65\%$ at $r=7\%$. Thus, a regular
withdrawal above the penalty threshold is a very bad strategy regardless the interest rate level. In this instance, the penalty
takes away the protected capital on a regular basis. Note that, the
penalty is applied in terms of the whole withdrawal amount, not just
on the exceeded part, see penalty function (\ref{eqn_penaltyP}). Thus a slight excess
over the penalty threshold can cause a large change in the price or
 in the fair fee, as observed in the second test.

The above two tests show that a regular static withdrawal is only
slightly beneficial at very low interest rate and only when the
withdrawal rate does not exceed the penalty threshold. In the next
section, it will be demonstrated by numerical results that an
optimal withdrawal is always beneficial regardless the
interest rate level and penalties.

\subsection{Optimal withdrawal -- super account}
Consider a GMAB for a super account with the annual ratchet and
assume that the policyholder can exercise an optimal withdrawal
strategy quarterly. For super account, any withdrawal will penalize
 the protected capital amount (benefit base) if the wealth account is below
the benefit base according to (\ref{guarantee_eq}) and (\ref{eqn_penaltyS}).

\begin{table}[!h]\begin{center}
\captionsetup{width=0.85\textwidth}
{\footnotesize{\begin{tabular*}{0.7\textwidth}{ccccccc} \toprule
\multirow{2}{*}{Interest rate, $r$} &\mbox{\;\;}& \multicolumn{2}{c}{$\sigma=10\%$ }   & \mbox{\;\;}& \multicolumn{2}{c}{$\sigma=20\%$} \\
& &  Fee (b.p) &   $\widetilde\varepsilon$ & & Fee (b.p) & $\widetilde\varepsilon$\\
 \midrule
$1\%$ & &370.7 & $10\%$& &1235   & $23.7\%$ \\
$2\%$ & &191.2 & $2.8\%$& & 700.1 & $9.89\%$\\
$3\%$ & &118.1 & $1.2\%$& & 478.8 & $4.54\%$\\
$4\%$ & &78.52 & $0.9\%$& & 355.5 & $2.48\%$\\
$5\%$ & & 54.47 & $1.1\%$& & 275.2 & $1.51\%$\\
$6\%$ & & 39.00 & $1.48\%$& & 218.8 & $1.16\%$\\
$7\%$ & & 28.38& $1.36\%$& & 176.9 & $1.03\%$\\
\bottomrule
\end{tabular*}
}} \caption{{\footnotesize{Fair fee $\alpha$ in bp (1 bp=0.01\%) as
a function of interest rate $r$ for the GMAB on a super account when withdrawals are optimal. The contract maturity is $T=10$ years. $\widetilde\varepsilon$ is the percentage difference between the fair fee in the case optimal withdrawal and the fair fee in the static case (no withdrawal) from Table \ref{tab_feeR1}. }}}\label{tab_feeSup}
\end{center}\end{table}

Table \ref{tab_feeSup} shows the fair fee for a super account as a
function of the interest rate at $\sigma=10\%$ and $\sigma=20\%$.
The columns under $\widetilde\varepsilon$ show an extra percentage value in the fee due to
optimal withdrawal when compared  to the static case of no withdrawal in Table \ref{tab_feeR1}. The results show, the extra fee is only about one or two percent for most cases, except   at the low interest rate
and high volatility. This extra fee
 due to optimal withdrawal is insignificant for the super account in most cases, mainly due to the heavy penalties applied.
 As will be shown in the next section, if the penalty is less severe as in the
 case of pension account, the extra fee becomes much more
 significant. If the penalty is completely removed, then numerical experiments show that the extra
 fee will be several times, e.g. $300\%$, larger than the static case,
 demonstrating the full value of the optimal strategy.

\subsection{Optimal withdrawal -- pension account}
Consider a GMAB for a pension account with the annual ratchet and
assume that the policyholder can exercise an optimal withdrawal
strategy quarterly. For a pension account, any withdrawal
 above a pre-defined withdrawal level $G_n$ will penalize the protected capital amount $A(t)$ if the wealth account $W(t)$ is below $A(t)$ according to (\ref{guarantee_eq})  and
(\ref{eqn_penaltyP}). Here, we set the annual withdrawal limit at $15\%$ of the
wealth account, i.e. $G_n=0.25\times 15\%=3.75\%$ is quarterly withdrawal threshold.

\begin{table}[!h]\begin{center}
\captionsetup{width=0.9\textwidth}
{\footnotesize{\begin{tabular*}{0.8\textwidth}{cccccccc} \toprule
\multirow{2}{*}{Interest rate, $r$} &\mbox{\;\;}& \multicolumn{2}{c}{$\sigma=10\%$ }   & \mbox{\;\;}& \multicolumn{3}{c}{$\sigma=20\%$} \\
& &  Fee (b.p) &   $\widetilde\varepsilon$ & & Fee (b.p) & $\widetilde\varepsilon$ & Fee-FD (b.p)\\
 \midrule
$1\%$& & 472.6 &  $39\%$ & & 1474 (1479) &  $46\%$ & 1466\\
$2\%$& & 227.7 &  $21\%$ & & 836.1 (836.3) &  $30\%$ & 833.7\\
$3\%$& & 135.4 &  $15\%$ & & 552.8 (553.6)&   $20\%$ & 551.7\\
$4\%$& & 88.15 &  $13\%$ & & 399.1 (399.7)&  $14\%$ & 398.6\\
$5\%$& & 60.24 &  $11\%$ & & 304.3 (304.7)&  $12\%$ & 304.0\\
$6\%$& & 42.58 &  $10\%$ & & 239.6 (239.9)&  $10\%$ & 239.4\\
$7\%$& & 30.63 &  $9\%$ & & 192.5 (192.8) &  $9\%$ & 192.4 \\
\bottomrule
\end{tabular*}
}}\caption{{\footnotesize{Fair fee $\alpha$ in bp (1 bp=0.01\%) as a
function of interest rate $r$ for the GMAB for a pension account when withdrawals are optimal. The contract maturity is $T=10$ years and quarterly withdrawal limit is $G_n=0.25\times 15\%$. $\widetilde\varepsilon$ is the percentage difference between the fair fee in the case optimal withdrawal and the fair fee in the static case (no withdrawal) from Table \ref{tab_feeR1}.  }}} \label{tab_feePension}
\end{center}\end{table}

Table \ref{tab_feePension} shows the fair fee of GMAB for a pension account
as a function of the interest rate $r$ when $\sigma=10\%$ and $20\%$.

The columns under $\widetilde\varepsilon$ show an extra percentage value in the fee due to
optimal withdrawal when compared  to the static case of no withdrawal in Table \ref{tab_feeR1}. The results show, the extra
fee ranges from about $9\%$ at the highest interest rate $r=7\%$ to
about $39\%$ at the lowest interest rate $r=1\%$. This extra fee is
much more significant than in the case of super account, see Table \ref{tab_feeSup}, apparently
due to reduced penalties. At $\sigma=20\%$, the extra fee ranges from
about $9\%$ at the highest interest rate  $r=7\%$ to about $46\%$ at
the lowest interest rate $r=1\%$. This extra fee is higher  than in the
case of lower volatility $\sigma=10\%$, in both percentage and
absolute terms.

Also, in Table \ref{tab_feePension}, the numbers in the parentheses next to the continuous fair fee values $\alpha$ are the
GHQC results for the discretely charged fair fee $\alpha_d= - \log (1-\widetilde{\alpha} dt_n)/dt_n$, where at the
end of each quarter $t_n$, the policyholder wealth account is charged a fee  proportional to
the account value $\widetilde{\alpha} dt_n W(t_n^-)$, see the wealth process (\ref{wealth_process_discrfee1}).
Results show only little difference between the continuous fee $\alpha$
and the discrete fee $\alpha_d$. On average, the relative difference
is $0.15\%$. Of course, at a higher frequency
of charging fee (e.g. a monthly fee), the difference will be even less.

\begin{figure}[!h]
\captionsetup{width=0.9\textwidth}
\begin{center}
\includegraphics[scale=0.57]{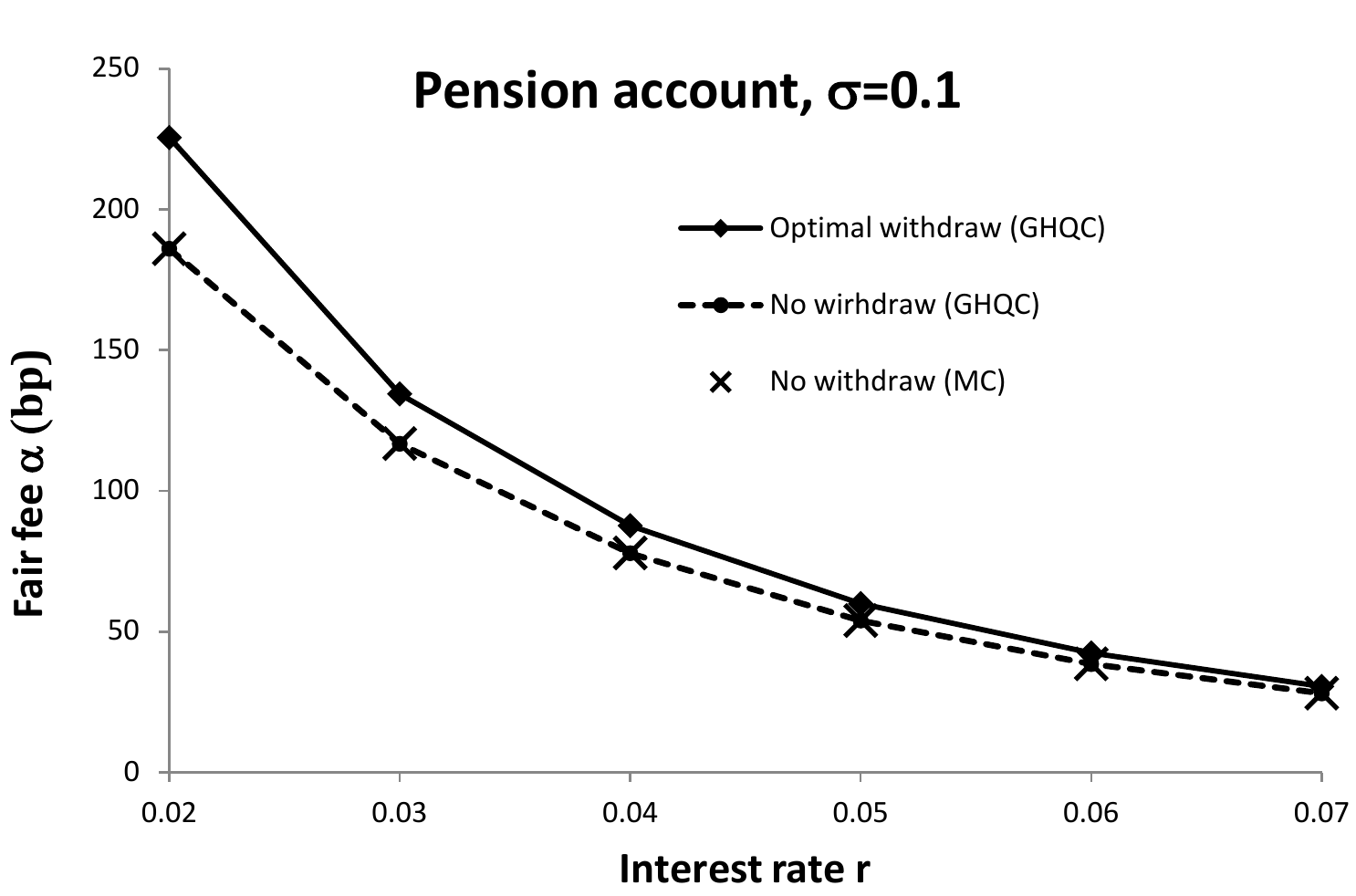}\includegraphics[scale=0.57]{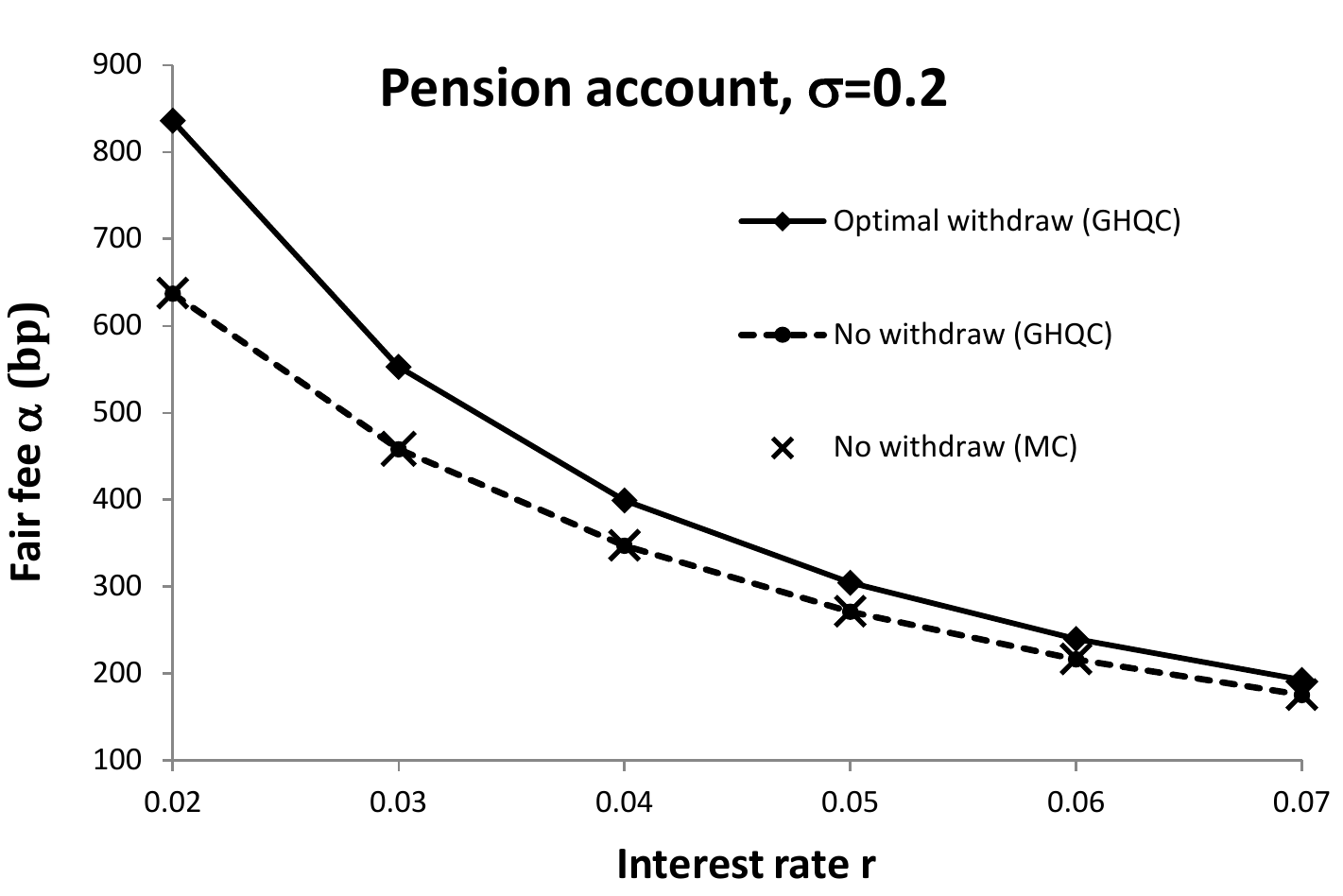}
\caption{{\footnotesize{Fair fee $\alpha$ in bp as a function of
the interest rate $r$ at
$\sigma=10\%$ and $\sigma=20\%$ for the GMAB on a pension account when withdrawals are optimal, in comparison with the static case where no withdrawal
is allowed. Values are taken from Table \ref{tab_feePension} and
Table \ref{tab_feeR1}, respectively. }}}\label{fig_feeP1}
\end{center}
\end{figure}

The last column in Table \ref{tab_feePension} shows the continuous
fee calculated using our finite difference PDE method. The
agreement between the GHQC (quadrature method) and finite difference PDE method is
very good. The average relative difference in the fair fees between the two
methods  is $0.20\%$. Figure \ref{fig_feeP1} 
show the curves of the fair fee for a pension account as a function of
interest rate $r$ using results from Table \ref{tab_feePension}, in
comparison with the static case (no withdrawal) from Table \ref{tab_feeR1}.

For some comparison, the market fees offered by
\cite{MLCproduct2014} are 1.75\% for a 10 year capital protection of
a ``balanced portfolio" and 0.95\% for a ``conservative growth
portfolio"; and fees offered by \cite{AMPNorthPDS2014} are 1.3\% for
a 10 year capital protection of a ``balanced strategy" portfolio and 0.95\%
for a ``moderately defensive strategy" portfolio. Though the values of
volatility are not known for these market portfolios, it seems that
market prices are significantly lower than the \emph{fair fee},
which is also observed in the literature before; e.g. see
\cite{milevsky2006financial}, \cite{bauer2008universal} and
\cite{ChenVetzalForsyth2008}.

\section{Conclusion}\label{conclusion_sec}
In this paper we have reviewed pricing VA
riders and presented a unified pricing approach via an
optimal stochastic control framework. We discussed different models and numerical procedures applicable in general to most of the VA riders with various contractual specifications.
To price these VA riders under the geometric Brownian motion model
for the risky asset, often assumed in practice, we have extended and generalized the direct
integration method based on the Gauss-Hermite quadrature, introduced earlier in \cite{LuoShevchenkoGMWB2015} for some specific and simpler product specifications.

As an example, we presented a numerical valuation of capital protection guarantees (GMAB riders), with specifications matching closely the real market products offered in Australia by e.g. \cite{MLCproduct2014} and \cite{AMPNorthPDS2014}.
Numerical valuation of this guarantee involves all the main numerical difficulties encountered in
pricing other VA riders, such as ratchets and optimal withdrawals. Numerical
results have been validated by MC and finite difference PDE methods and can serve as a benchmark for practitioners
and researchers developing numerical pricing of VA riders. As expected,  we observed that the
extra fee that has to be charged to counter the optimal policyholder
behavior is most significant at lower interest rate and higher
volatility levels, and it is very sensitive to the penalty withdrawal
threshold.

As we have already discussed in Section \ref{remark_withdrawalstrategy_sec}, the fee based on the optimal policyholder withdrawal is the worst
case scenario for the issuer, i.e. if the guarantee is hedged then
this  fee will  \emph{ensure} no losses for the issuer (in other
words full protection against policyholder strategy and market
uncertainty). If the issuer hedges
continuously but investors deviate from the optimal strategy, then
the issuer will receive a guaranteed profit.
Any strategy different from the optimal is sup-optimal and will
lead to smaller fair  fees. Of course the strategy optimal in this
sense is not related to the policyholder circumstances. The
policyholder may act optimally with respect to his preferences and
circumstances which may be different from the optimal strategy maximizing losses for the policy issuer.
Life-cycle modelling can be undertaken to analyze and estimate sub-optimality of policyholder behavior. However, development of secondary markets for insurance products may expose the policy issuers to some significant risk if a fee for the guarantees is not charged to cover the worst case scenario.

It is important to note that the guarantee could be written on more that one asset (several mutual funds). In this case it is still common for practitioners to use a single-asset proxi model to calculate the price and hedging parameters. Obviously such approach has significant drawbacks (e.g. the sum of geometric Brownian motions is not a geometric Brownian motion). PDE and direct integration methods are not practical in high-dimensions and thus one has to rely on the MC methods to treat multi-asset case accurately. In the case of static withdrawal, it is not difficult to consider full multi-asset model and calculate the price using standard MC as in \cite{ng2013pricing}. However, in the case of optimal withdrawal strategies, numerical valuation in the multi-asset case will require development of regression type MC solving backward recursion for  processes affected by the withdrawals. One could apply \emph{control randomization} methods extending Least-Squares MC developed in \cite{Kharroubi2014},  but the accuracy robustness of this method for pricing  VA riders have not been studied yet.

The specification details of VA riders typically vary across different companies and are  difficult to extract and compare from the very long product specification documents. Moreover, results for specific GMxB riders presented in academic literature often refer to different specifications. As a result, cross-validation and benchmark research studies
are rare. Given that numerical solutions used for pricing of VA riders are complex, it is important that these solutions are properly tested and validated.
Moreover, new products are appearing in VA market regularly with increasing complexity that raises an important question, as discussed in  \cite{carlin2011obfuscation} and mentioned in \cite{bauer2015behavior}, whether new complex products are designed to suite the policyholder needs better or introduced for the purpose of \emph{obfuscation}.

\section{Acknowledgement} This
research was supported by the CSIRO-Monash Superannuation Research
Cluster, a collaboration among CSIRO, Monash University, Griffith
University, the University of Western Australia, the University of
Warwick, and stakeholders of the retirement system in the interest
of better outcomes for all. This research was also partially supported under the
Australian Research Council's Discovery Projects funding scheme (project number:
DP160103489). We would like to thank Man Chung Fung for many constructive comments.

{\footnotesize{
\bibliographystyle{chicago} 
\bibliography{bibliography}
}}

\end{document}